\documentclass[preprint,sort&compress,3p]{elsarticle}

\usepackage{url}
\usepackage{amsmath}
\usepackage{amsfonts}
\usepackage{bm}

\usepackage{graphicx}
\usepackage{multirow}
\usepackage{multicol}
\usepackage{float}
\usepackage{subfigure}
\usepackage{hyperref}

\usepackage{lineno}

\newcommand{\numu}[0]{$\nu_\mu$}

\newcommand{\nue}[0]{$\nu_e$}

\newcommand{\thetaatm}[0]{$\theta_{23}$}

\newcommand{\thetasol}[0]{$\theta_{12}$}

\newcommand{\thetareac}[0]{$\theta_{13}$}
\newcommand{\deltacp}[0]{$\delta_{CP}$}

\newcommand{\pizero}[0]{$\pi^{0}$}

\newcommand{\numi}[0]{NuMI}

\newcommand{\annem}[0]{$\textrm{ANN}_{e\mu}$}
\newcommand{\annnc}[0]{$\textrm{ANN}_{NC}$}

\begin{document}

\title{CHIPS Event Reconstruction and Design Optimisation}
\author[lan]{A. Blake}
\author[ucl]{S. Germani}
\author[mad]{Y. B. Pan}
\author[ucl]{A. J. Perch}
\author[ucl]{M. M. Pf\"{u}tzner}
\author[ucl]{J. Thomas}
\author[ucl]{L. H. Whitehead\fnref{f1}}
\fntext[f1]{Now at CERN, Geneva, Switzerland}
\address[ucl]{Department of Physics and Astronomy, University College London, Gower Street, London, WC1E  6BT, United Kingdom}
\address[mad]{Department of Physics, University of Wisconsin, Madison, WI 53706, United States of America}
\address[lan]{Physics Department, Lancaster University, Lancaster, LA1 4YB, United Kingdom}

\begin{abstract}

The CHIPS experiment will comprise a $10\,$kton water Cherenkov detector in an open mine pit in northern Minnesota, USA. The detector has been simulated using a full GEANT4 simulation and a series of event reconstruction algorithms have been developed to exploit the charge and time information from all of the PMTs. A comparison of simulated CCQE \numu{} and \nue{} interactions using 10\,inch and 3\,inch PMTs is presented, alongside a comparison of 10\% and 6\% photocathode coverage for 3\,inch PMTs. The studies demonstrate that the required selection efficiency and purity of charged-current \nue{} interactions can be achieved using a photocathode coverage of 6\% with 3\,inch PMTs. Finally, a dedicated $\pi^{0}$ fitter is shown to successfully reconstruct a clean sample of $\pi^{0}$ mesons despite the low 6\% photocathode coverage with 3\,inch PMTs.

\end{abstract}


\maketitle


\section{Introduction}

It has been conclusively shown over the last 20 years that neutrinos undergo oscillations from one flavour to another, and the field has now entered a precision measurement era. The oscillations arise because the three weak interaction eigenstates are not the same as the three mass eigenstates but are formed from a linear combination, as defined by the PMNS matrix~\cite{mns1962,ponte1968,ponte1969}. The PMNS matrix is commonly expressed in terms of four parameters: three mixing angles \thetasol{}, \thetaatm{} and \thetareac{}; and the $CP$-violating phase \deltacp{}. The oscillations themselves are driven by the mass-squared differences between the mass eigenstates $\Delta m^{2}_{ij}\equiv m^{2}_{i} - m^{2}_{j}$ with $i > j$, giving two independent mass-squared splittings.

With the measurement of a non-zero value of \thetareac{} by reactor antineutrino experiments\cite{dayaBay2012,reno2012,doubleChooz2012}, the main focus of the neutrino oscillation community has moved to resolving the mass hierarchy ambiguity and measuring the value of the CP-violating phase, \deltacp{}. Long-baseline neutrino oscillation experiments can probe the mass hierarchy and \deltacp{} by looking for \numu{} $\rightarrow$\nue{} transitions. The \nue{} appearance probability, taking into account the MSW matter effect \cite{wolf1978,ms1986}, expanded to second order in $\alpha\equiv\Delta m^2_{21}/\Delta m^2_{32}$ can be written as \cite{nueAppForm}

\begin{align}\label{int:eq:appProb}
P(\nu_\mu\rightarrow\nu_e) \approx \sin^{2}\theta_{23}\sin^{2}&2\theta_{13} \frac{\sin^{2}\Delta(1-A)}{(1-A)^2} +\alpha\tilde{J}\cos(\Delta\pm\delta_{CP})\frac{\sin\Delta A}{A}\frac{\sin\Delta(1-A)}{(1-A)} \nonumber \\
&+\alpha^2\cos^{2}\theta_{23}\sin^{2}2\theta_{12}\frac{\sin^{2}\Delta A}{A^{2}},
\end{align}
with $\Delta = \Delta m^2 _{31} L/4E$, $\tilde{J}=\cos\theta_{13}\sin2\theta_{13}\sin2\theta_{12}\sin2\theta_{23}$, and $A = 2\sqrt{2}G_{F}n_{e}E/\Delta m^{2}_{31}$, where $G_{F}$ is the Fermi weak coupling constant and $n_{e}$ is the electron density. The sign of $A$ is positive for neutrinos and negative for antineutrinos. Similarly, the sign in front of \deltacp{} in the second term is positive for neutrinos and negative for antineutrinos. Equation \ref{int:eq:appProb} shows that the \numu{}$\rightarrow$\nue{} probability has sensitivity to the mass hierarchy through the matter effect parameter $A$ and to the $CP$-violating phase through the second term.

Water Cherenkov (WC) neutrino detectors infer the presence of a neutrino interaction by measuring the Cherenkov light \cite{cherenkov1939} emitted by charged particles resulting from the interaction. Cherenkov radiation is emitted by all electrically charged particles travelling faster than the local speed of light in a dielectric medium. The charged particles emit a cone of radiation with an opening angle $\theta_C$ defined as
\begin{equation}\label{int:eq:thetac}
\cos\theta_C = \frac{1}{\beta n},
\end{equation}
where $n$ is the refractive index of the material and $\beta$ is the particle speed divided by the speed of light in the vacuum. Inside a detector, this cone of light projects onto the wall and forms a ring shape. 

The study of neutrino oscillations generally requires clean samples of charged-current (CC) interactions for different neutrino flavours. Water Cherenkov detectors are well suited to distinguishing CC \numu{} and \nue{} interactions, particularly at low neutrino energies when the track multiplicity is lower. These CC interactions proceed by the exchange of a $W$ boson and are of the form $\nu_l + X \rightarrow l^{-} + X'$, where the flavour of the final-state lepton, $l$, is used to identify the flavour of the neutrino. Electrons initiate an electromagnetic cascade meaning that the produced Cherenkov radiation is actually the sum of the individual Cherenkov cones from all of the electrons and positrons in the cascade, and hence the ring of light detected is wider and fuzzier than the corresponding ring from a muon. There are three main types of CC interaction, ranging from those with minimal hadronic activity to those with a high energy transfer to the hadronic system: quasi-elastic (QE), resonant (RES) and deep inelastic scattering (DIS). 

The primary background to the CC \nue{} appearance search, once the CC \numu{} interactions have been identified, comes from the production of \pizero{} mesons in neutral-current (NC) interactions: $\nu + X \rightarrow \nu + X' + \pi^{0}$. The \pizero{} meson decays into a pair of photons with a probability of 98.8\%~\cite{pdg}, and can mimic an electron signal in two ways: firstly, when the \pizero{} has a high boost the decay photons are emitted almost colinearly and produce closely overlapping rings of light on the detector wall; and secondly, if one photon is not detected then the single remaining photon ring is indistinguishable from that of an electron, since the photon will also initiate an electromagnetic shower. The first of these two possibilities is reducible in the event that the two closely overlapping rings can be reconstructed and shown to give a better fit than a single electron ring.


\section{The CHIPS Experiment}

The CHerenkov detectors In mine PitS (CHIPS) experiment plans to deploy a WC neutrino detector into a large body of fresh water, such as a flooded mine pit. The main goal of the project is to construct a 10\,kton water Cherenkov detector at a dramatically reduced cost compared to previous detectors through novel design choices. The water surrounding the detector will provide the structural support and sufficient overburden, and means that no large-scale civil engineering is required, as would be necessary for building a detector in an underground cavern. A demonstration of this ability would pave the way towards future affordable mega-ton scale neutrino detectors.  

CHIPS will only study neutrino interactions from accelerator neutrino beams. This means that the design requirements with regard to photodetectors are different from previous and existing large WC neutrino detectors such as Super-Kamiokande \cite{superKNIM}. These experiments have wide physics programs including low energy physics arising from solar neutrinos and proton decay searches. The energy threshold of water Cherenkov detectors is primarily dictated by the photocathode coverage, and the cost of photodetectors, such as photomultiplier tubes (PMTs), is a major driver to the total cost of the experiment. Considering only accelerator beam neutrinos enables the photocathode coverage to be reduced since the energy threshold can be much larger than for solar neutrino studies, for example. 

The CHIPS project began with the deployment of CHIPS-M, a small prototype detector 3.3\,m tall and 3\,m in diameter, into a water-filled, dis-used mine pit in northern Minnesota in the summer of 2014. The mine pit is located approximately 700\,km from the NuMI \cite{numiBeam} beam target at Fermilab and 7\,mrad off-axis. CHIPS-M was then subsequently refurbished and redeployed in the same location in the summer of 2015. CHIPS-M was instrumented with five 10\,inch PMTs and after refurbishment, an additional 32 3\,inch PMTs. The next step will be the building of a $10\,$kton module called CHIPS-10, which is planned for deployment in the summer of 2018. This document details simulation and event reconstruction studies to benchmark the performance of different detector geometry options.


\section{Simulation} \label{sec:simulation}

A GEANT4 \cite{geant4} simulation has been written to simulate the CHIPS detectors. It was originally based on the WCSim package \cite{wcsim} developed for the proposed T2K 2km and LBNE WC detectors, but has been overhauled and almost entirely rewritten to better suit the requirements for CHIPS. The main design goal of the simulation was to produce a versatile package to allow many different geometry options to be considered without the need for recompilation.

The simulation builds an n-sided, regular polygonal prism detector geometry consisting of the top and bottom end caps separated by the main barrel region. The dimensions of the geometry are stored in a configuration file and are loaded at run time. Each of the sides of the n-sided barrel form a separate \emph{region} and each of the caps can be divided into regions defined by the angle subtended. A concentric veto volume is included in order to study the efficiency of rejecting cosmic-ray muons and the effect of cosmic-ray muons overlapping with a beam neutrino interaction.

\begin{figure}
\centering
\includegraphics[scale=0.5]{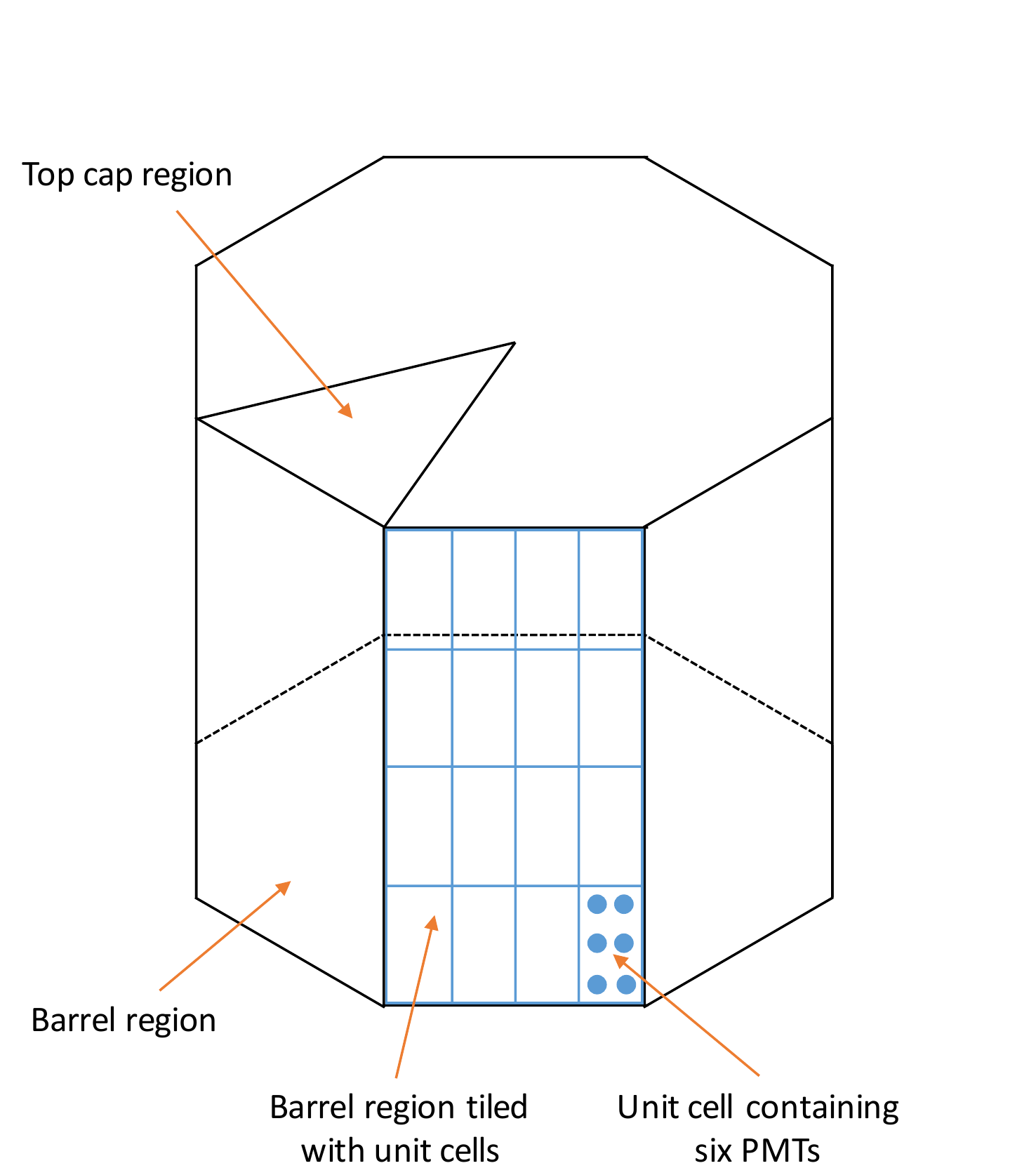}
\caption{\label{sim:fig:detGeo}An illustration of an eight-sided inner detector showing regions defined on the cap and barrel. One of the barrel regions has been tiled with unit cells, and an example unit cell is shown to contain six PMTs. The region on the top cap is defined by the opening angle.}
\end{figure}

The base unit of the geometry is known as the \emph{unit cell}. Each region of the detector can have a uniquely defined unit cell that contains a pattern of PMTs. PMTs can be placed such that they face outwards in order to look at the veto region as well as inwards to observe the main detector volume. This pattern can consist of any number of PMTs of different sizes and types that are defined in a second configuration file. The geometry is then built by tiling each region with its corresponding unit cell, and the unit cells are scaled in size to produce the required level of photocathode coverage for the region. The unit cell represents what would be a plane of PMTs in the real detector. Figure \ref{sim:fig:detGeo} shows a cartoon of the inner detector with example regions and unit cells displayed.

A simple dynode chain simulation was written to convert the number of photons incident on the PMT photocathode returned by GEANT4 into a digitised number of photoelectrons. The simulation assumes a total gain of $1\times10^{7}$ across ten dynode chains, the relative gains of which are taken from the base of the 10\,inch Hamamatsu PMT developed for the IceCube experiment~\cite{icecubePMT}, the PMT used in the first deployment of the CHIPS-M prototype. The simulation also accounts for non-linearity at high numbers of incident photons resulting from the electron cloud in the dynode chain effectively shielding some of the accelerating electric field.

\subsection{CHIPS-10 Geometries}\label{sim:sec:geom}

The CHIPS-10 detector is simulated as a 20-sided prism with a height of $20\,$m and a $12.5\,$m radius. The two different sizes of PMT considered here are $10\,$inch and $3\,$inch. The properties of the 3\,inch PMT, other than the actual diameter, were set to be identical to the 10\,inch PMT in order to be able to perform a direct comparison of the effect of the PMT size. Both 10\,inch and 3\,inch tubes were assumed to have high quantum efficiency, peaking at 34\% at $380\,$nm. These parameters can be trivially changed in a configuration file once an actual PMT has been selected for the experiment. The three geometry options considered are:
\begin{itemize}
  \item Uniform 10\% coverage of 10\,inch PMTs
  \item Uniform 10\% coverage of 3\,inch PMTs
  \item Uniform 6\% coverage of 3\,inch PMTs.
\end{itemize}
In order to obtain the same photocathode coverage using 3\,inch PMTs and 10\,inch PMTs, the 3\,inch PMT geometry has a factor 11 more PMTs based on the ratio of the areas of the PMTs.

The attenuation length of the water as a function of wavelength is parametrised from $200-800\,$nm using data from measurements of pure water~\cite{waterAbsQuickenden,waterAbsPope,waterAbsBuiteveld} and uniformly scaled down to account for impurities such that it peaks at $50\,$m at $405\,$nm. A recent dedicated  study showed that the attenuation length attainable using just filtration and UV-sterilisation of the water from the mine pit is up to 100\,m \cite{waterPaper}, meaning this could be a slightly conservative choice.

\subsection{Event Generation}
\begin{figure}
\centering
\includegraphics[scale=0.4]{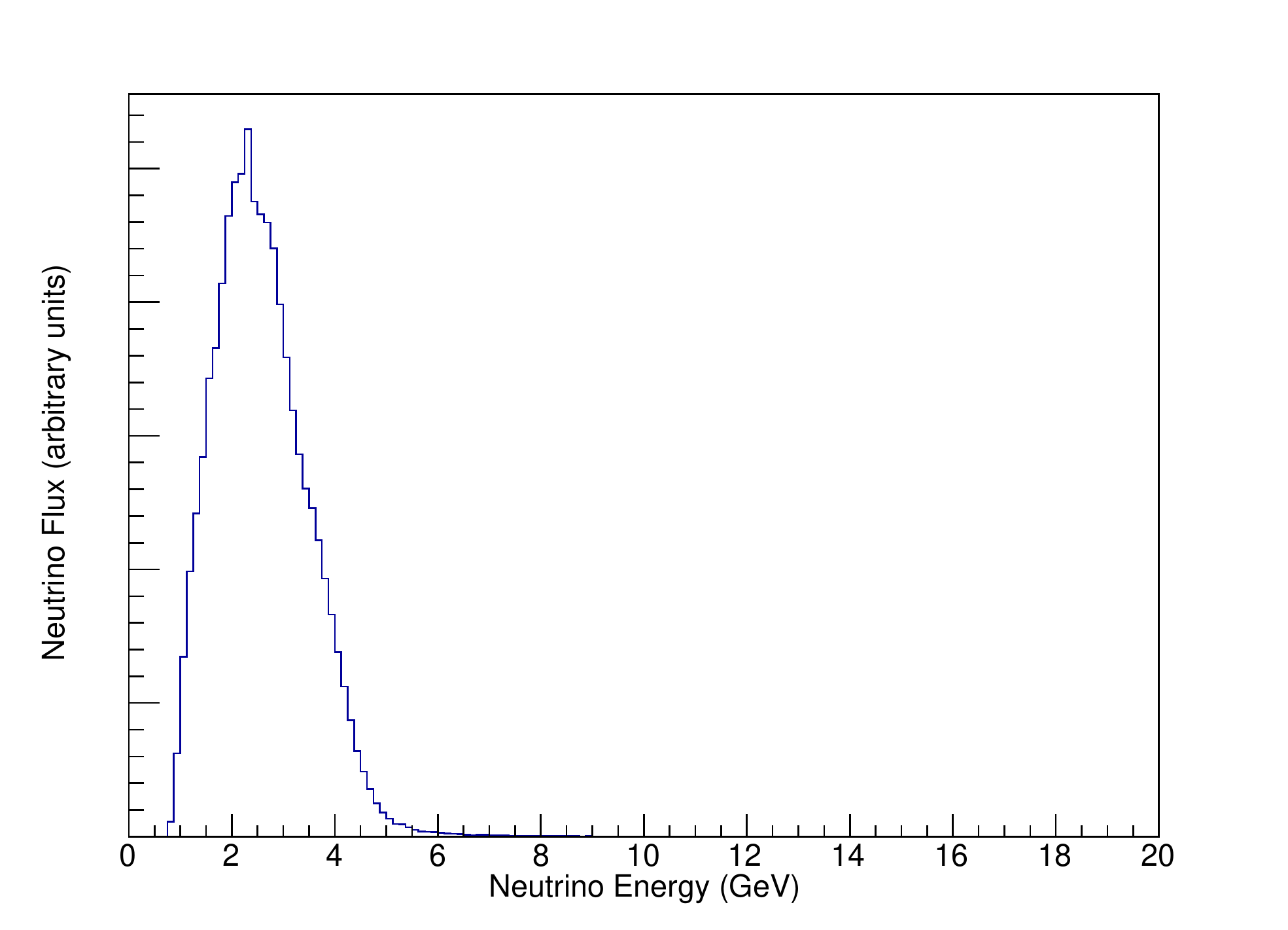}
\caption{\label{sim:fig:flux}The simulated NuMI muon neutrino flux at a distance of 700\,km and 7\,mrad off-axis.}
\end{figure}

The generation of beam neutrino events is performed by the GENIE~\cite{GENIE} package using as input the predicted \numi{} \cite{numiBeam} beam spectrum at the CHIPS detector location, 700\,km from the target and 7\,mrad off-axis. The predicted spectrum is that of the \numu{} flux, shown in Fig. \ref{sim:fig:flux}, which is then oscillated to provide the \nue{} flux using the values $\Delta m^2_{32} = 2.39\times10^{-3}$eV$^2$, $\sin^2\theta_{23} = 0.43$, $\sin^22\theta_{13}=0.0945$, $\delta_{CP} = 0$, and the solar scale oscillation terms are neglected. Cosmic-ray muons are generated with the CRY~\cite{CRY} package, producing as output a series of cosmic-ray muons with their positions spread over a square kilometre at the Earth's surface surrounding the detector. The interactions from both generators are converted into the same file format and used as input to the detector simulation.

The beam interaction or cosmic-ray muon files are passed into the detector simulation to produce events. It is also possible to use a combination of particles with beam and cosmic origins in order to produce beam events with cosmic-ray muons overlaid to study how the reconstruction software can cope with overlapping events. The beam events are assigned a random time within the \numi{} beam spill window of approximately $10\,\mu$s. The full simulated events are then produced with the GEANT4 simulation described previously.


\section{Reconstruction}
The first steps of the reconstruction are employed in order to seed the tracks for the full likelihood fit that forms the bulk of the reconstruction method. The seeding algorithms are used to provide the main fitter with a good starting point from which to begin the minimisation routines, and to allow it to efficiently find the optimal set of parameters that describe the interaction.

\subsection{Track seeding}\label{rec:sec:seeding}
The first stage in seeding the tracks is to try to find clearly separated energy deposits in space and time. The digitised hits from each PMT are sorted into time order and sliced up when gaps are found in time between sequential hits. Each time slice is then considered in turn to look for distinct spatial collections of PMT hits. A charge cut is applied at this stage in order to prevent low charge hits joining up the different regions\footnote{These hits are only removed from the seeding process, all hits are considered in the full likelihood fit.}. Hits are then clustered together beginning with the highest charged hit, and clustering continues until no further clusters can be formed. The slicing algorithm then returns a series of slices that consist of a collection of associated PMT hits.

A series of vertexing algorithms are applied individually to the slices in order to produce estimates of the interaction vertex position, vertex time and track direction. The slices are then passed through a Hough transform algorithm to further refine the track direction values, and to look for other sub-dominant directions that could indicate that the energy deposit was caused by multiple particles. The output from the Hough transform, and hence the seeding algorithms, is a list of seeds where each seed consists of a vertex position, vertex time and track direction. The particle energy is not estimated in the seeding algorithms.

The outputs from the Hough transform for all slices are sorted by their peak height in the Hough-transform space, where a large height corresponds to a stronger seed. The number of tracks, $n$, defined before the fit, are then set with initial parameters from the $n$ best seeds.

\subsubsection{Cosmic-ray muon track seeding}
The detector geometry includes a veto volume in order to determine whether a beam interaction also contains a cosmic-ray muon track. Traditionally veto volumes are used to define some dead time to the detector in order to avoid contamination of beam interactions with the cosmic-ray muons. However, in order to minimise the detector dead time, an algorithm has been developed to exploit the information from the veto region to tag and seed the cosmic-ray muon tracks. This process requires the identification of the cosmic-ray muon entry and exit point, and as such, only works for through-going cosmic-ray muons.

The algorithm begins by attempting to cluster the veto PMT hits. If two (or more) clusters can be formed, the largest two are taken and the entry point is assigned as the position of the highest charge hit in the upper cluster, and the exit point to the highest charge hit in the lower one. If only a single cluster is found, then the highest charge hits in the top and bottom halves of the detector are taken as the entry and exit points. The track vertex is then simply assigned as the entry point, the track direction from the vector joining the entry and exit points, and the time from the hit time of the hit with the highest charge at the entry point.

\begin{figure}
  \centering
  \includegraphics[scale=0.4]{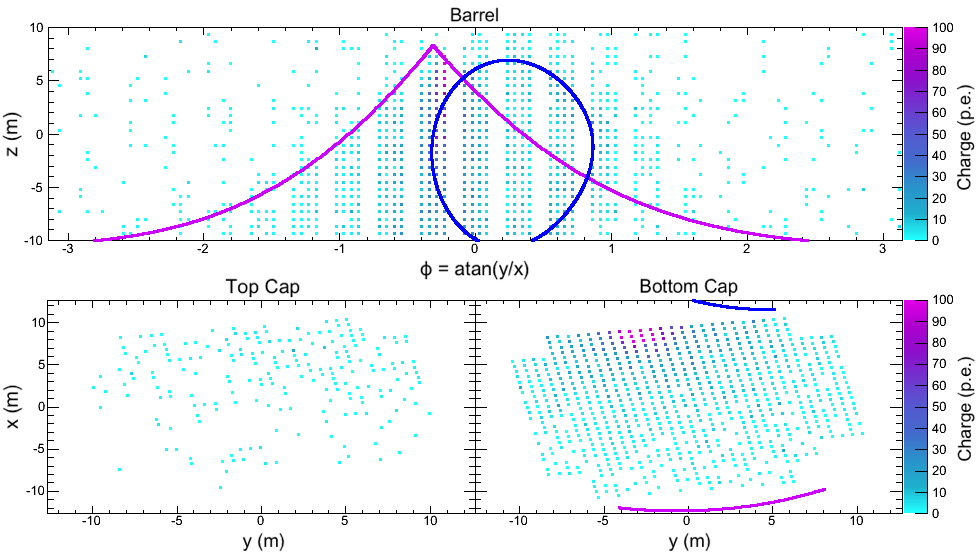}
  \includegraphics[scale=0.5]{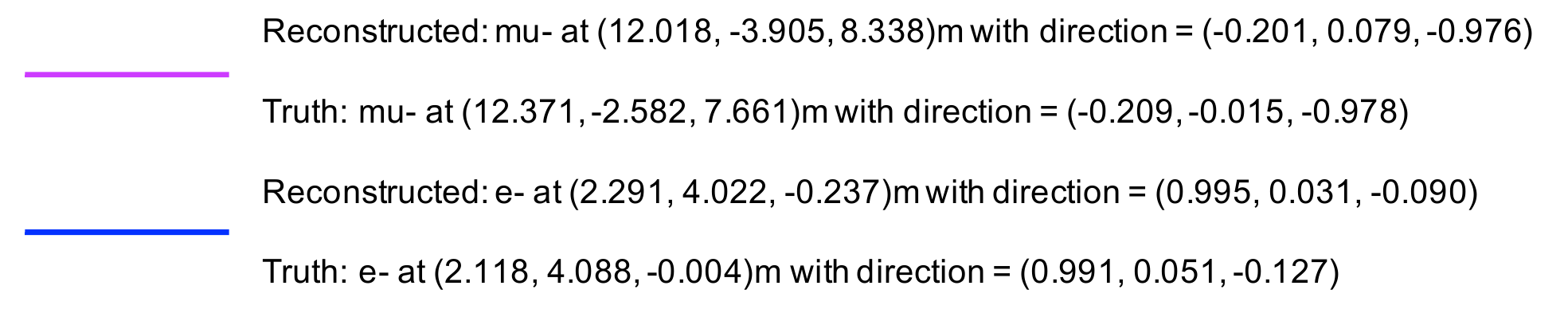}
  \caption{\label{rec:fig:cosmicSeed}An event display of a simulated CCQE \nue{} interaction with an overlaid cosmic-ray muon. The top panel shows the unrolled barrel section of the detector and the top and bottom caps are shown in the bottom left and bottom right panels, respectively. The colour shows each inner PMT that received an energy deposit and represents the collected charge, limited to a maximum of 100\,p.e. for clarity. The two rings show projections of the true tracks onto the detector wall. The figure was produced immediately after the seeding algorithm, meaning no further reconstruction was performed, and hence no attempt to reconstruct the particle energies was made.}
\end{figure}

If a good cosmic-ray muon seed is obtained it is used to form a shadowed region inside the inner detector. All those inner detector hits that fall within the Cherenkov cone defined by the cosmic-ray muon seed are masked out of the standard seeding algorithm outlined previously, such that the beam interaction can also be accurately seeded. Figure \ref{rec:fig:cosmicSeed} shows an event display of an example output from seeding a cosmic-ray muon overlaid on to a beam \nue{} CCQE interaction.

\subsection{Likelihood track-fitting method}\label{rec:sec:likelihood}
The reconstruction algorithms for CHIPS are based on calculating the likelihood to observe energy deposits of a given charge and time for a particular particle track hypothesis. The algorithms were written from scratch, with the charge prediction based upon the likelihood method developed for the MiniBooNE experiment~\cite{pattersonThesis,minibooneReco}. The method is relavitely simple in theory: given a hypothesised track or tracks, the number of photoelectrons and time at which the first photoelectron is detected for each PMT are predicted. The measured hit charges and times can then be compared to these predictions in order to calculate the likelihood that the given track hypothesis was responsible for the measured signals. The parameters that define the track hypothesis are then varied until the negative logarithm of the likelihood is minimised, giving the best-fit track hypothesis to the measurements.

The method uses information from all PMTs, whether there was a registered charge deposit or not. The method can be extended in order to perform multiple track fits and this is a vital part of identifying the $\pi^{0}$ decays that form a background to the \nue{} appearance search. The method is also very modular such that only one of the components depends on the detector geometry. This means it is easy to test different detector geometry configurations without having to recalculate many of the parts that go into the likelihood.

The basic form of the likelihood consists of two components:
\begin{equation}\label{rec:eq:basicLikelihood}
 \mathcal{L}(\bm{x}) = \mathcal{L}_{\textrm{unhit}}(\bm{x}) \mathcal{L}_{\textrm{hit}}(\bm{x}) = \prod_{\textrm{unhit}} P_{\textrm{unhit}}(\bm{x})  \prod_{\textrm{hit}} P_{\textrm{charge}}(\bm{x}) P_{\textrm{time}}(\bm{x}),
\end{equation}
where $\bm{x}$ is the track hypothesis. The first term calculates the likelihood that a given track $\bm{x}$ will not predict a hit in those PMTs without registered hits. The second term calculates the likelihood that the same track hypothesis \emph{will} cause a hit PMT to register a given charge at a given time. The track hypothesis $\bm{x}$ consists of the following parameters:
\begin{itemize}
  \item The track vertex position $(x_0,y_0,z_0)$ and time $t_0$.
  \item The initial track direction $(d_x,d_y,d_z)$.
  \item The initial kinetic energy of the particle.
  \item The particle type - muon, electron or photon.
\end{itemize}

The charge prediction calculation is identical for both the PMTs with and without registered charge deposits, and is discussed in Section \ref{rec:sec:charge}. The PMTs with registered hits also must have a hit time prediction which is detailed in Section \ref{rec:sec:time}. 

\subsubsection{Charge component}\label{rec:sec:charge}
There are three main steps in order to calculate the charge component of the likelihood: 
\begin{enumerate}
\item Predict the mean number of photons incident on the PMT photocathode.
\item Convert the number of photons into the output digitised charge from the PMT.
\item Calculate the likelihood to see the measured energy deposit given the predicted charge.
\end{enumerate}

The formalism to perform the first step was described in detail by the MiniBooNE collaboration \cite{pattersonThesis,minibooneReco}. Adaptations to the method have been made to account for the differences between the experiments, such as the cylindrical geometry in CHIPS as opposed to the spherical symmetry of the MiniBooNE detector, and the lack of scintillation light from water compared to mineral oil. Considerable detail of the adaptations and the specific implementation of the algorithms is given in Ref. \cite{andyThesis}.

An extended track, where light is emitted at each point along the length, can be considered as a series of point sources. Each of these sources produces a certain number of photons, $\mu_i$ that reach the PMT. These contributions are summed over the track to give the total number of photons as:
\begin{equation}\label{eq:reco:initialMu}
\mu = \sum_{i}\mu_{i} = \sum_{i} \Phi_{i} T(R_{i}) \epsilon(\psi_{i}) \frac{\Omega(R_{i})}{4\pi}.
\end{equation}
The components of $\mu_{i}$ are as follows: $\Phi_{i}$ is the number of photons emitted in step $i$ in the direction of the PMT; $T(R_{i})$ is the transmission function and parametrises the probability that a photon will survive after traversing a distance $R_{i}$, the distance from the point source $i$ to the PMT; $\epsilon(\psi_{i})$ is the angular efficiency of the PMT as a function of the angle of the incident photon, where $\psi_{i} = 0$ along the PMT normal direction; and $\frac{\Omega(R_{i})}{4\pi}$ is the fraction of the total solid angle subtended by the PMT as viewed from the point source $i$. 

\begin{figure}
\centering
\includegraphics[scale=0.5]{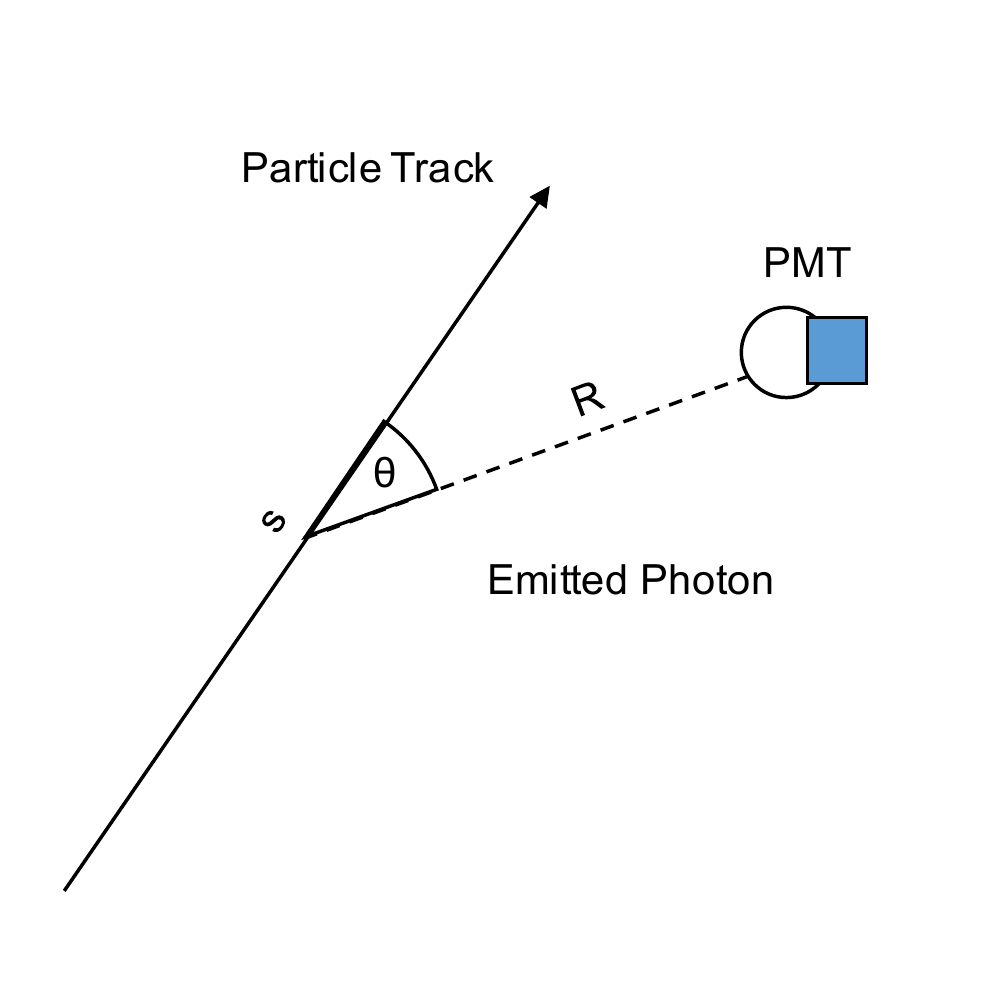}
\caption{\label{reco:fig:coordSyst}The definition of the distance along the track $s$, the angle of the emitted photon $\theta$ and the distance travelled by the photon $R$.}
\end{figure}

To first order, only $\Phi_{i}$ depends on the particle type and energy and can be written as
\begin{equation}\label{eq:reco:phiExpanded}
	\Phi_{i} = \Phi(E)\rho(E,s_{i})g(E,s_{i}, \cos{\theta(s_{i})}),
\end{equation}
where $s_i$ is the distance along the track and $\theta(s_i)$ is the angle of the emitted photon from the track direction, illustrated in Fig. \ref{reco:fig:coordSyst}. $\Phi(E)$ is the total number of photons emitted by a particle of energy $E$. The function $\rho(E,s_{i})$ gives the fraction of all photons emitted in step $i$, and $g(E,s_{i}, \cos{\theta(s_{i})})$ gives the number of photons from step $i$ emitted in the angular range $\cos\theta \rightarrow \cos\theta + d(\cos\theta)$, and normalised such that the sum over $\cos\theta$, multiplying by the bin width, for a given $s$ is equal to one. The energy dependence of the emission profiles $\rho(E,s_{i})$ and $g(E,s_{i}, \cos{\theta(s_{i})})$ is handled by binning in energy. This means that for each particle type and energy bin, two emission profiles are created, $\rho(s_{i})$ and $g(s_{i}, \cos{\theta(s_{i})})$, using large samples of Monte Carlo simulated events.

Combining Eqs. \ref{eq:reco:initialMu} and \ref{eq:reco:phiExpanded} and noting that since the track vertex, track direction and PMT position are fixed for a given hypothesis, both $\psi_{i}$ and $R_{i}$ depend solely on $s_i$, gives
\begin{equation}\label{eq:reco:secondMu}
\mu = \Phi(E)\sum_{i} T(s_{i}) \epsilon(s_{i}) \frac{\Omega(s_{i})}{4\pi}\rho(E,s_{i})g(E,s_{i}, \cos{\theta(s_{i}))}.
\end{equation}

In calculating the likelihood, Eq. \ref{eq:reco:secondMu} must be evaluated for each PMT and for each iteration of the track hypothesis, which would prove to be very computationally intensive. In order to combat this, as described in Refs.~\cite{pattersonThesis,minibooneReco}, Eq. \ref{eq:reco:secondMu} can be rewritten as
\begin{equation}\label{eq:reco:thirdMu}
\mu = \Phi(E)\sum_{i}J(s_{i})\rho(E,s_{i})g(E,s_{i}, \cos{\theta(s_{i}))}.
\end{equation}
with $J(s_{i}) = T(s_{i}) \epsilon(s_{i}) \frac{\Omega(s_{i})}{4\pi}$. Since $J(s)$ is a slowly varying function, it can be expanded quadradically in terms of $s$: $J(s) \approx j_0 + j_1 s + j_2 s^2$. The coefficients of the expansion are calculated using three values of $s$ and solved by matrix inversion, rather than actually performing a fit. The three values are chosen to cover a large portion of the track length, but differ depending on whether the track is contained inside the detector. For contained tracks, the points are at: $10\,$cm, the distance at which $75\%$ of photons have been emitted and half this distance. For exiting tracks, the three points are at: $10\,$cm, and $40\%$ and $80\%$ of the distance to the exiting point. Writing $\xi(E,s_{i},\cos{\theta(s_i)}) = \rho(E,s_{i})g(E,s_{i}, \cos{\theta(s_{i}))}$, Eq. \ref{eq:reco:thirdMu} becomes
\begin{equation}\label{eq:reco:finalMu}
\mu = \Phi(E)\left(j_0 \sum_{i}\xi(E,s_{i},\cos{\theta(s_i)}) + j_1\sum_{i} s_i \xi(E,s_{i},\cos{\theta(s_i)}) + j_2\sum_{i} s^2_i \xi(E,s_{i},\cos{\theta(s_i)}) \right).
\end{equation}
The summations in Eq. \ref{eq:reco:finalMu} depend only on the emission profiles and can hence be calculated in advance for each particle type and stored in a look-up table as a function of the track energy, distance from the vertex to the PMT $R_0$, and the angle between the direction from the vertex to the PMT and the particle direction $\cos\theta_0$. This means that the computationally intensive part is performed once and then the values are just used by the fitter from the look-up table. It is important that there is a smooth behaviour as a function of energy from the look-up table, hence cubic splines are used to interpolate between energy bins. For a given $\left(R_0,\cos\theta_0\right)$ bin the spline is loaded and evaluated at the desired energy.

The next step is to convert the predicted number of photons $\mu$ into a number of photoelectrons $\mu_{pe}$. This is performed by applying the quantum efficiency of the PMT, the values of which are parametrised as a function of photon wavelength in the simulation. The quantum efficiency is averaged over the wavelengths of light expected at the PMT, since the wavelength spectrum changes as a function of the distance travelled by the photons.

The final step is to calculate the likelihood of the measured PMT charge given the predicted number of photoelectrons for each PMT $p$. This is done by considering Poisson statistics with $\lambda_p = \mu_{pe,p}$ for each PMT by comparing to the measured charge $q_p$ in the following way:
\begin{equation}\label{rec:eq:chargePoissLikelihood}
  \mathcal{L}_{\textrm{charge}} = \prod_{\textrm{hit}}P_{\textrm{charge}} = \prod_{p}\frac{\lambda_{p}^{q_p}e^{-\lambda_p}}{q_p!},
\end{equation}
or in $-2\ln\mathcal{L}$ form, noting also that the Gamma function $\Gamma(n+1)=n!$ is used to replace the factorial since $q_p$ is not constrained to be an integer,
\begin{equation}\label{rec:eq:chargePoissLogLikelihood2}
  -2\ln\mathcal{L}_{\textrm{charge}} = -2\sum_{p}q_p\ln\lambda_{p} - \lambda_p - \ln \Gamma(q_p+ 1).
\end{equation}
In the case that the PMT had no measured charge deposit, i.e. $q_p=0$, the contribution to the likelihood simplifies to $2\lambda_p$.

\subsubsection{Time component}\label{rec:sec:time}
The treatment of the time component of the fit differs substantially from the method used by MiniBooNE, and is hence described here in detail. A further fully detailed description can be found in Ref. \cite{andyThesis}. The method described here tries to predict the time at which each PMT will receive a hit by calculating the trajectories of the photons emitted along the length of the track. The large size of CHIPS-10 compared to MiniBooNE means that there is a much larger spread in PMT hit times within an event, and hence the time of the hits provides a lot of information. The time difference between the first and last hits, not including scattered light, is of the order of $50\,$ns for CHIPS-10, which when compared with typical PMT timing resolutions of 1 or 2$\,$ns, shows the potential power of the timing information.

The predicted hit time $t_{PMT}$ of a photon incident on a PMT at position $\bm{x_{PMT}}$ emitted by a particle with a vertex four-vector $(\bm{x_0},t_0)$ and momentum unit vector $\bm{\hat{p}}$, travelling through a medium with refractive index $n$, can be written as the sum of three terms: the vertex time, the time taken until the photon emission and the travel time of the photon. This gives

\begin{equation}\label{eq:reco:pmtHitTime}
t_{PMT} = t_0 + \frac{s}{v} + \left( \frac{n}{c}\times|\bm{x_{PMT}} - (\bm{x_0} + s\bm{\hat{p}})| \right),
\end{equation}
where $s$ is the distance along the track at which the photon was emitted and $v$ is the speed of the charged particle. The effective speeds of the muons and electromagnetic showers were found to be $0.94c$ and $0.81c$ from the simulation, respectively. These effective speeds account for the fact that the emission of photons does not occur along a perfectly straight line. It is clear that if the particle only emits photons at a single angle then there is only one value of $s$ that will cause a hit on a given PMT. In reality, however, the photons are emitted at a range of angles, and this range is much larger for electrons than muons since the photons are emitted by all of the electrons and positrons in the electromagnetic shower. There are hence different combinations of $s$ and the distance travelled by the photon that can cause a hit on a given PMT. 

The time prediction proceeds by stepping along the length of the track in units of $25\,$cm. At each step $\Delta s$, the goal is to calculate the time that a hit would occur on each PMT, and a corresponding weight that describes how likely it is to be hit at that time, given the distribution of light emitted. The time of the hit is simply given by Eq. \ref{eq:reco:pmtHitTime}. The weight must be calculated from emission profiles, similar to those described in the calculation of the charge prediction. The profiles $\rho_t(E,s_i)$ and $g_t(E,s_i,\cos\theta_i)$, where $\rho_t(E,s_{i})$ gives the fraction of all photons emitted in step $i$, and $g_t(E,s_{i}, \cos{\theta(s_{i})})$ gives the number of photons from step $i$ emitted in the angular range $\cos\theta \rightarrow \cos\theta + d(\cos\theta)$, are normalised differently to their charge equivalents such that the weight, equal to the fraction of the total number of photons emitted towards the PMT in the $\Delta s$ step, is given by:

\begin{equation}
w_{i} = \rho_{t}(s_{i})g_{t}(s_{i}, \cos{\theta_{i}}) \Delta s_{i} \Delta{\cos{\theta_{i}}}.
\end{equation}

Hence, for each PMT, a distribution of hit times and their corresponding weights is obtained after stepping along the entire length of the track. The predicted hit time is then taken as the weighted mean of this array of points calculated in the following way:

\begin{equation}
\bar{t}_{PMT} = \left(\displaystyle\sum\limits_{i}{t_{i}w_{i}}\right)/\left(\displaystyle\sum\limits_{i}{w_{i}}\right)
\end{equation}

The weighted RMS also provides important information for calculating the likelihood. This width $\sigma_{source}$, describes the intrinsic uncertainty in the hit time due to the emission of photons by the charged particle, and is defined as:

\begin{equation}
\sigma^2_{source} = \left(\displaystyle\sum\limits_{i}(t_{i} - \bar{t}_{PMT})^{2}w_{i}\right) / \left(\displaystyle\sum\limits_{i} w_{i}\right).
\end{equation}
In the event that only one step of $\Delta s$ provides a predicted hit time, the width is set to be $\sigma_{source} = 0.1\,$ns in order for later calculations to succeed.

The mean and RMS of the predicted time, along with the probability weights, are interpolated as a function of energy. In particular, Catmull-Rom \cite{catmullRom} splines were used such that both the values and their derivatives vary smoothly between the energy bins. They are also very computationally fast as they rely only on the nearest four points to construct the spline.

There is one further complication in the calculation when considering PMTs hit by multiple photons; the hit time recorded by the PMT is the time of the first hit, not the mean of all the photons. Assuming that the distribution of photon arrival times is Gaussian with mean $\bar{t}_{PMT}$ and width $\sigma_{source}$, then the larger the number of photons considered, the earlier the first hit time should fall within that distribution. Order statistics gives the probability density functions $P_{k}(t)$ for the $k^{\textrm{th}}$ smallest value obtained from $n$ values drawn from a probability distribution $f(t)$. In this case the minimum value corresponds to $k=1$ such that
\begin{equation}\label{eq:reco:orderStats}
  P(t_{min}) = n \left( 1 - F(t) \right)^{(n-1)} f(t),
\end{equation}
where $t_{min}$ is the earliest time, $F(t)$ is the cumulative distribution function of $f(t)$, and $n$ is the number of photons detected. For a Gaussian $f(t)$ Eq. \ref{eq:reco:orderStats} becomes:
\begin{equation}\label{eq:reco:finalOrderStats}
  P(t_{min}) = \frac{n}{{2^{n}}\sigma\sqrt{2}}\left(1 - \textrm{erf}{\left(\frac{t - \mu}{\sqrt{2}\sigma}\right)}\right)^{n-1}\exp{\left(-\frac{(t-\mu)^{2}}{2\sigma^{2}}\right)}.
\end{equation}

The probability for the PMT to be hit at the predicted time is equal to the overlap between the $P(t_{min})$ distribution and a Gaussian distribution, $f_{PMT}$, with a mean set to the measured PMT hit time and the width equal to the PMT time resolution. Mathematically, this overlap area is described as
\begin{equation}
P_{\textrm{time}} = \int^{\infty}_{-\infty} \textrm{min}(P(t),f_{PMT}(t))dt
\end{equation}
It is faster to calculate this value by finding the points where the functions cross, and using the exact integrals of the two functions to compute the area.

\subsection{Single Track Fits}
In the case of a single track, the best track seed is taken from the seeding algorithms described in Section \ref{rec:sec:seeding}. The minimisation covers a complicated parameter space with non-trivial correlations. In particular there is a three-way correlation between the track vertex position along the track direction, the vertex time and the energy, such that a movement in the vertex position can be compensated by altering the vertex time and energy to give a similar fit result. The minimisation routine proceeds in the following steps:
\begin{enumerate}
 \item Fit the time using the time component of the fit only.
 \item Fit the energy using both charge and time components.
 \item Fit the energy, track vertex along the track direction, and the time, and then re-fit the track direction. This step is then repeated once more.
 \item Fit all track parameters using charge and time components.
 \item Fit the energy and track vertex along the track direction using charge only.
 \item Fit the energy with charge and time components.
 \item Fit the time with the charge and time components.
\end{enumerate}

\begin{figure}
  \centering
  \includegraphics[scale=0.3]{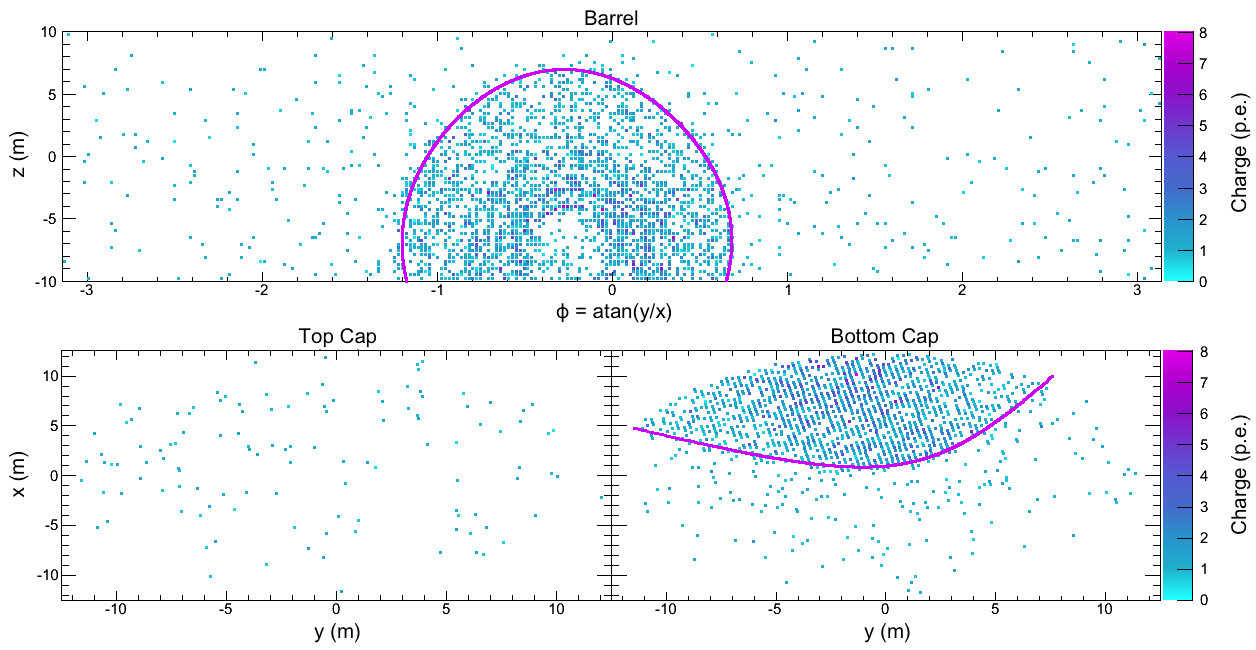} 
  \includegraphics[scale=0.3]{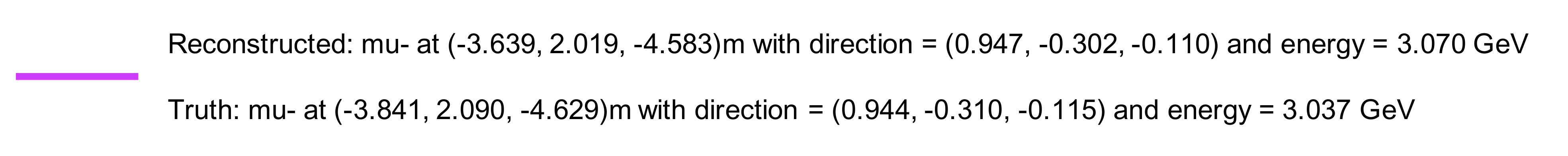}
  \includegraphics[scale=0.3]{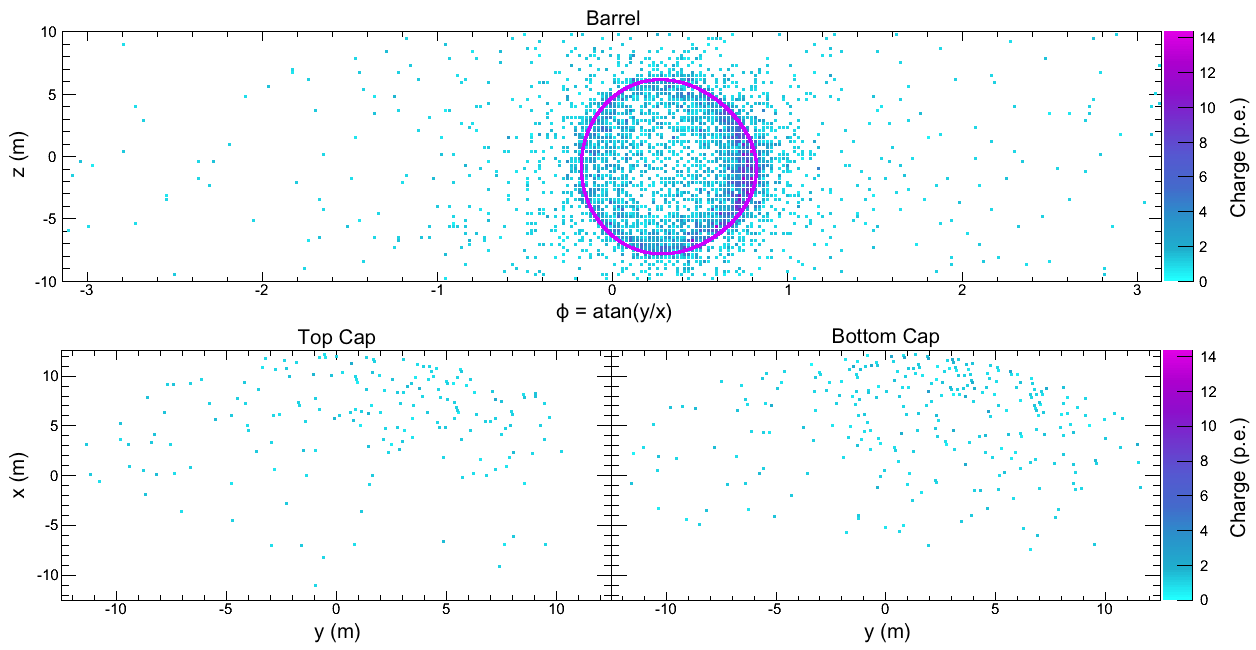}
  \includegraphics[scale=0.3]{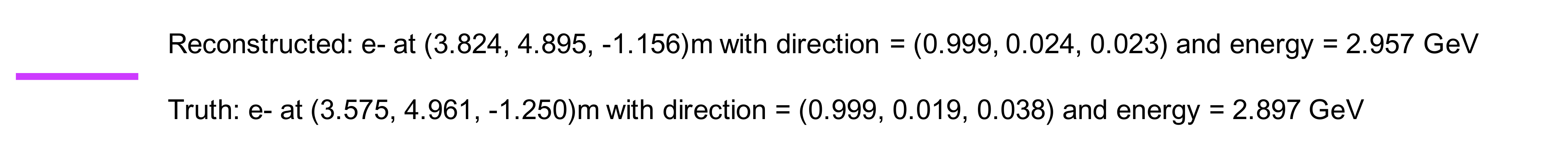}
  \caption{\label{rec:fig:evDisplays}The top (bottom) event display shows the result of a single track fit to a simulated CCQE \numu{} (\nue{}) interaction under the muon (electron) hypothesis using a 6\% coverage of 3\,inch PMTs. The magenta ring shows the best-fit from the reconstruction algorithms.}
\end{figure}

The track parameters are varied during the minimisation process until the minimum combined likelihood from the charge and time components of the fit is obtained. The values of the parameters at the minimum value of the negative log-likelihood are the best estimate of the true values. Figure \ref{rec:fig:evDisplays} shows example single track fits of a CCQE \numu{} interaction under the muon hypothesis (top), and a CCQE \nue{} interaction under the electron hypothesis (bottom), showing good agreement between the reconstructed and true parameter values.

\subsection{Multiple Track Fits}
The reconstruction method is relatively straightforward to extend to a multiple track hypothesis. The predicted charge for each PMT simply becomes the sum of the predicted charges from each of the $j$ tracks:
\begin{equation}
\mu_{total} = \sum_{j} \mu_{j}
\end{equation}
For the time component, the contribution to the time likelihood for each PMT from each of the tracks, $\mathcal{L}^{t}_{j}$, is weighted by the predicted charge in the PMT:
\begin{equation}
\mathcal{L}^{t}_{total} = \frac{\sum_{j} \mu_{j}\mathcal{L}^{t}_{j}}{\sum_{j}\mu_{j}} 
\end{equation}

Seeding multiple track events can be challenging, and Section \ref{rec:sec:seeding} outlined the standard track seeding method and the cosmic-ray muon track seeding. A special procedure is used in the case of seeding tracks for the $\pi^{0}$ fits.

\subsubsection{$\pi^{0}$  Fits}
One of the main challenges in identifying \nue{} interactions is to efficiently and accurately separate them from background events containing $\pi^0$ mesons. Unless good seeds are found for the two decay photon tracks it is hard to prevent the fit from collapsing down into an effective single track fit. The procedure begins by performing a single-track electron fit in order to provide a baseline likelihood to which the two-track fit can be compared. Two algorithms are then applied in order to obtain seeds for the two $\pi^{0}$ decay photon tracks.

Firstly, the two best track seeds are then taken from the output of the standard seeding algorithm described in Section \ref{rec:sec:seeding}. A common vertex is formed by extrapolating the tracks backwards to a point of closest approach, and the initial energies of the tracks are set to be in the ratio of the height of the peaks in the Hough transform under the condition of the invariant mass of the $\pi^{0}$. This vertex, the track directions and energies are then used to create two photon hypothesis tracks to represent the $\pi^{0}$ and the likelihood is calculated. If this likelihood is better than the single electron likelihood then this seed is passed to the fitter, otherwise a second attempt to seed a $\pi^{0}$ is made using a method similar to the procedure used by MiniBooNE \cite{pattersonThesis,minibooneReco}, outlined below.

The PMT hit positions are projected onto a plane containing the single electron vertex position and perpendicular to the electron direction. The major and minor axes of the covariance ellipse of the hits are formed. Rotations are then performed on the electron track to represent the first of the two photons: 0, $\pm\frac{\pi}{20}$, $\pm$0.45 and $\pm\frac{\pi}{5}$ radians about the major axis; and $\pm\frac{\pi}{20}$ radians about the minor axis. The second photon track direction is obtained through a grid search between $\pm\frac{\pi}{8}$ in both the $\theta$ and $\phi$ directions with respect to the first photon direction. The range of the grid search was chosen to concentrate on the small separations where the Hough transform does not perform as well. The two configurations giving the best likelihood when the second photon has greater than or less than half the energy of the first photon are passed on to the fitter. Both of these are then minimised in the fitter, and the best likelihood is returned.


\section{Comparison of PMT options}\label{sec:pmtComp}
A study was performed to benchmark the reconstruction software performance for the three geometry options discussed in Sec. \ref{sim:sec:geom}. These options are 10\,inch PMTs at 10\% coverage, and 3\,inch PMTs at both 10\% and 6\% coverage. Samples of CCQE \numu{} and CCQE \nue{} interactions were produced using the simulation, with energy spectra following those expected from the \numi{} beam. The events were reconstructed using a single track fit with the track hypothesis set to match the flavour of the leading charged lepton. Following the reconstruction, a preselection was applied to select events that should be well reconstructed: 
\begin{itemize} 
\item Track does not exit the detector
\item Number of hit PMTs greater than 50
\item Vertex position at least 1\,m from the detector wall
\item Track energy in the range $550\,$MeV to $4950\,$MeV (this is the range covered by the generated emission profiles).
\end{itemize} 

\begin{figure}
  \begin{center}
    \begin{tabular}{cc}
      \includegraphics[scale=0.28]{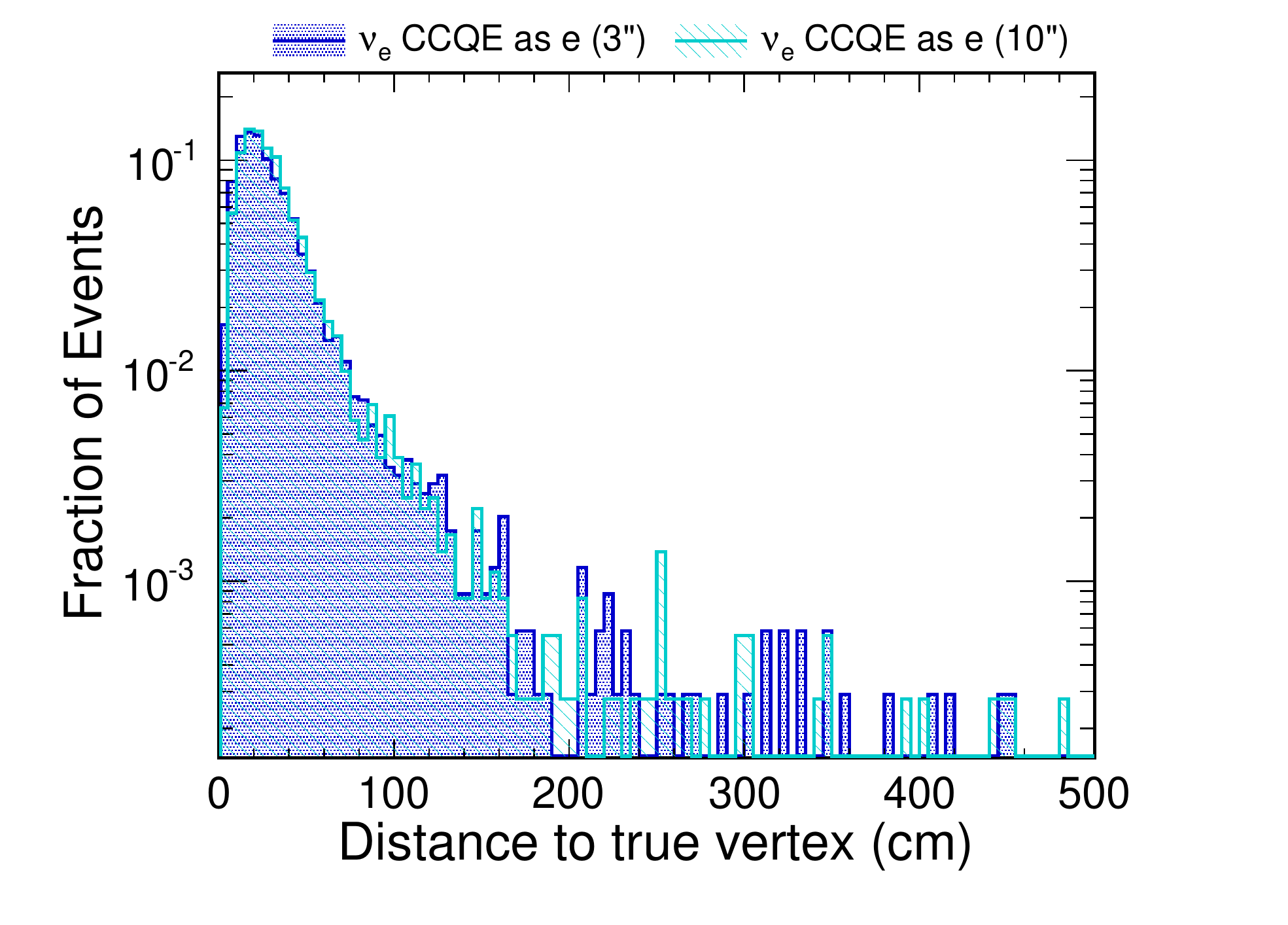} &
      \includegraphics[scale=0.28]{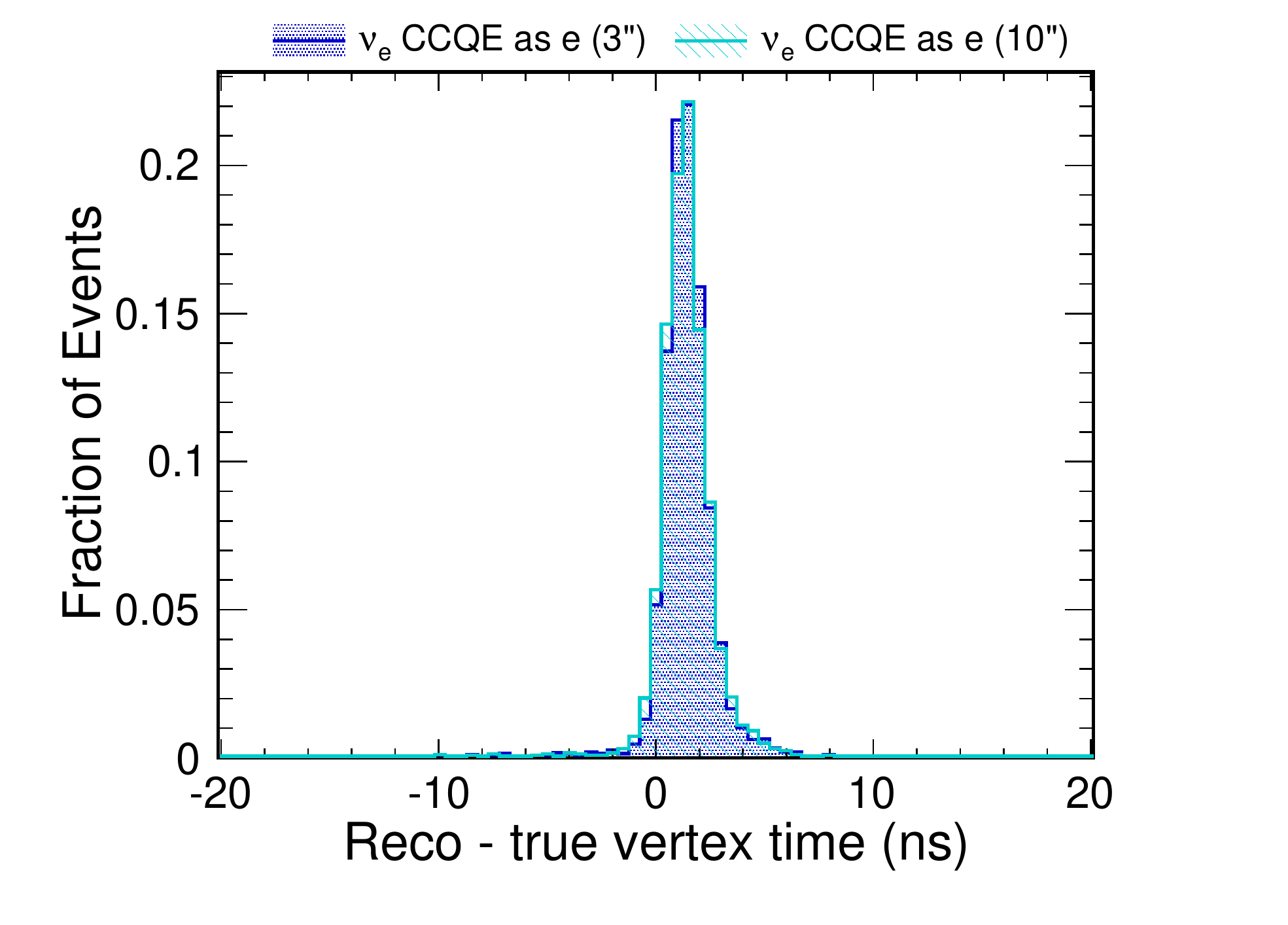} \\
      \includegraphics[scale=0.28]{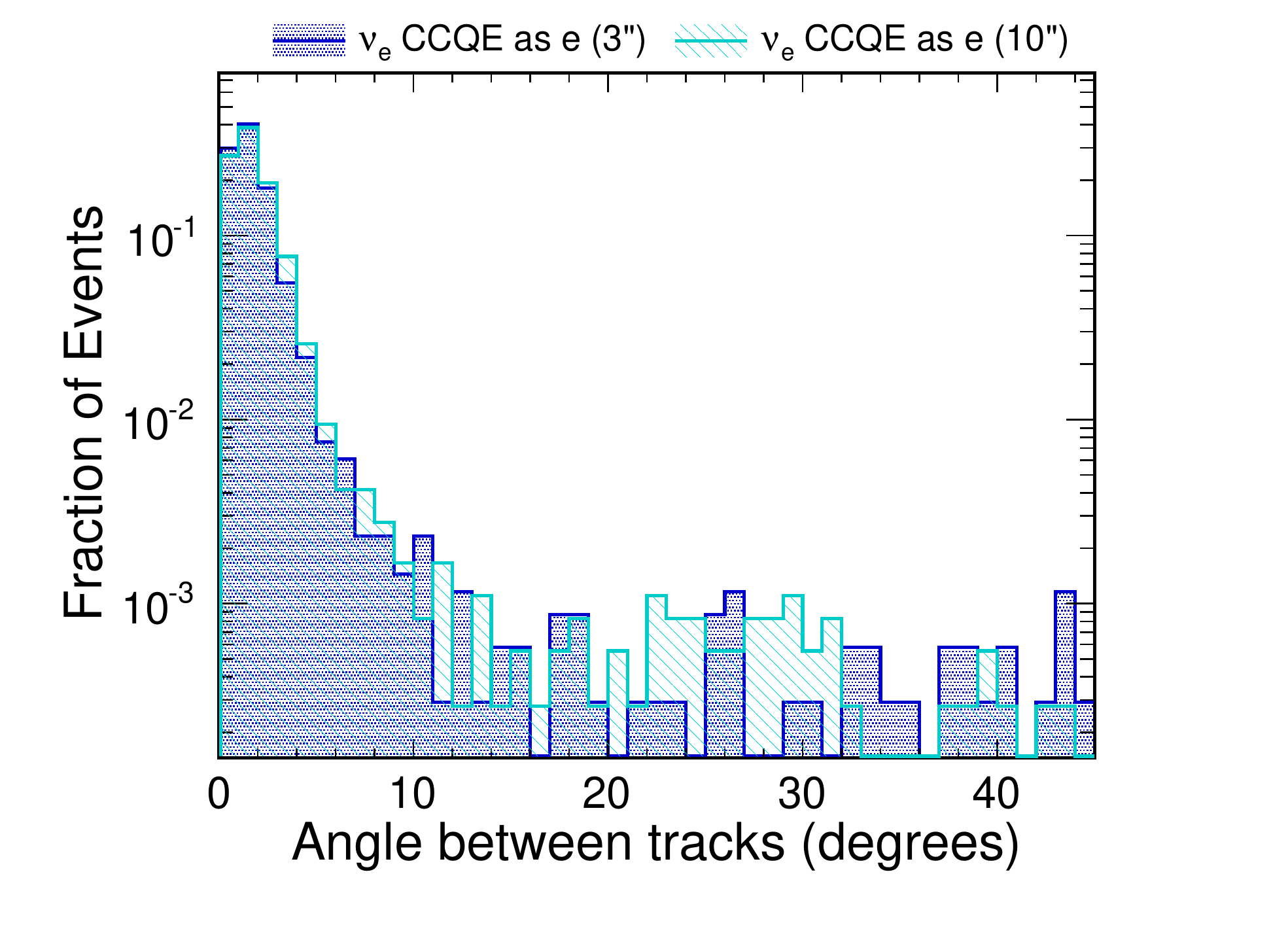} &
      \includegraphics[scale=0.28]{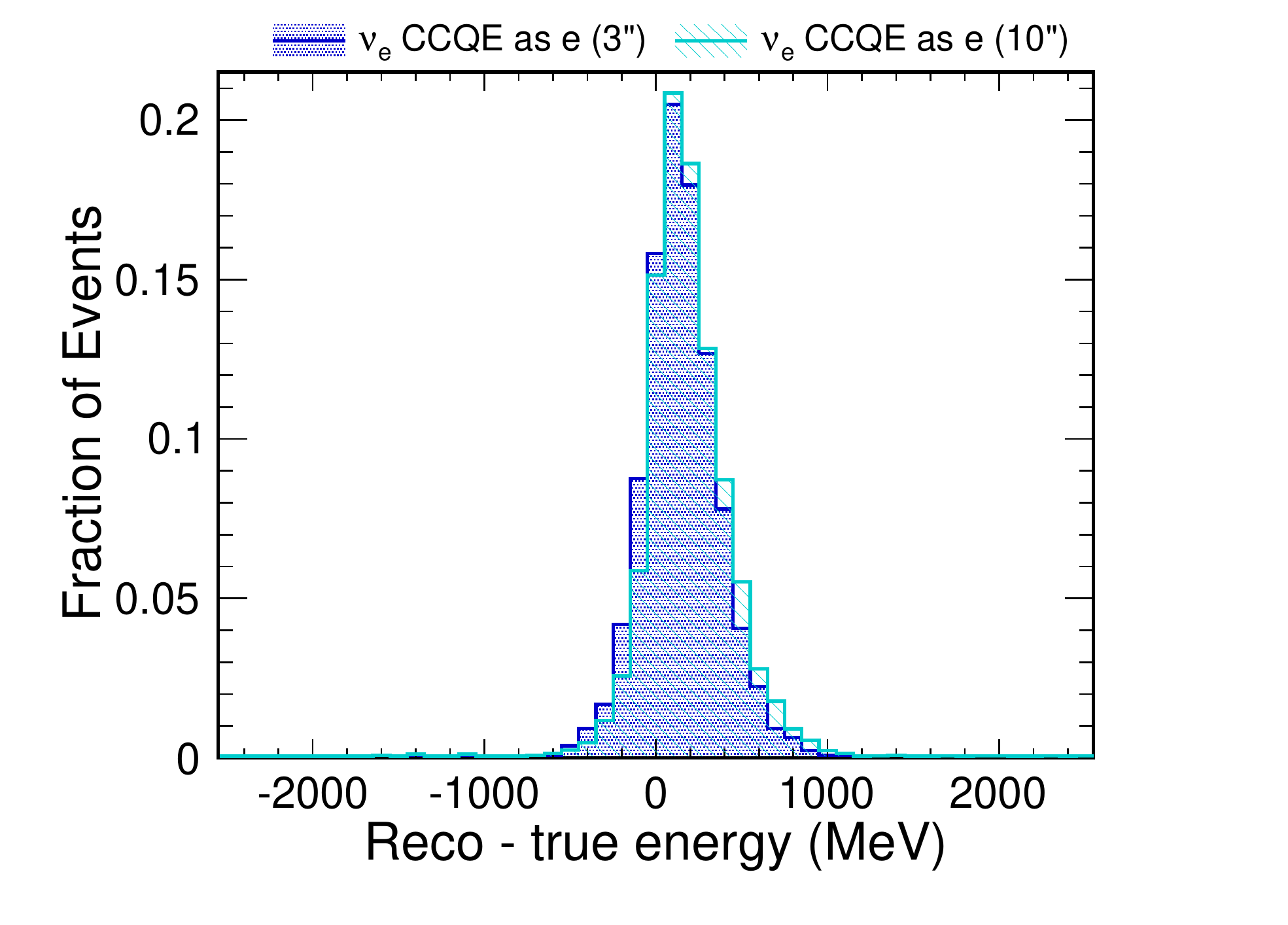} 
    \end{tabular}
  \caption{\label{perf:fig:10vs3NuE}A comparison of the reconstruction performance for a single track fit to CCQE \nue{} interactions for 10\,inch and 3\,inch PMTs for a photocathode coverage of 10\%. The distributions show (clockwise from top left): the distance of the reconstructed vertex from the simulated vertex, the difference between the reconstructed and simulated values for the vertex time, the difference between the reconstructed and simulated values for the electron energy and the angle between the reconstructed and simulated track directions.}
  \end{center}
\end{figure}

\begin{figure}
  \begin{center}
    \begin{tabular}{cc}
      \includegraphics[scale=0.28]{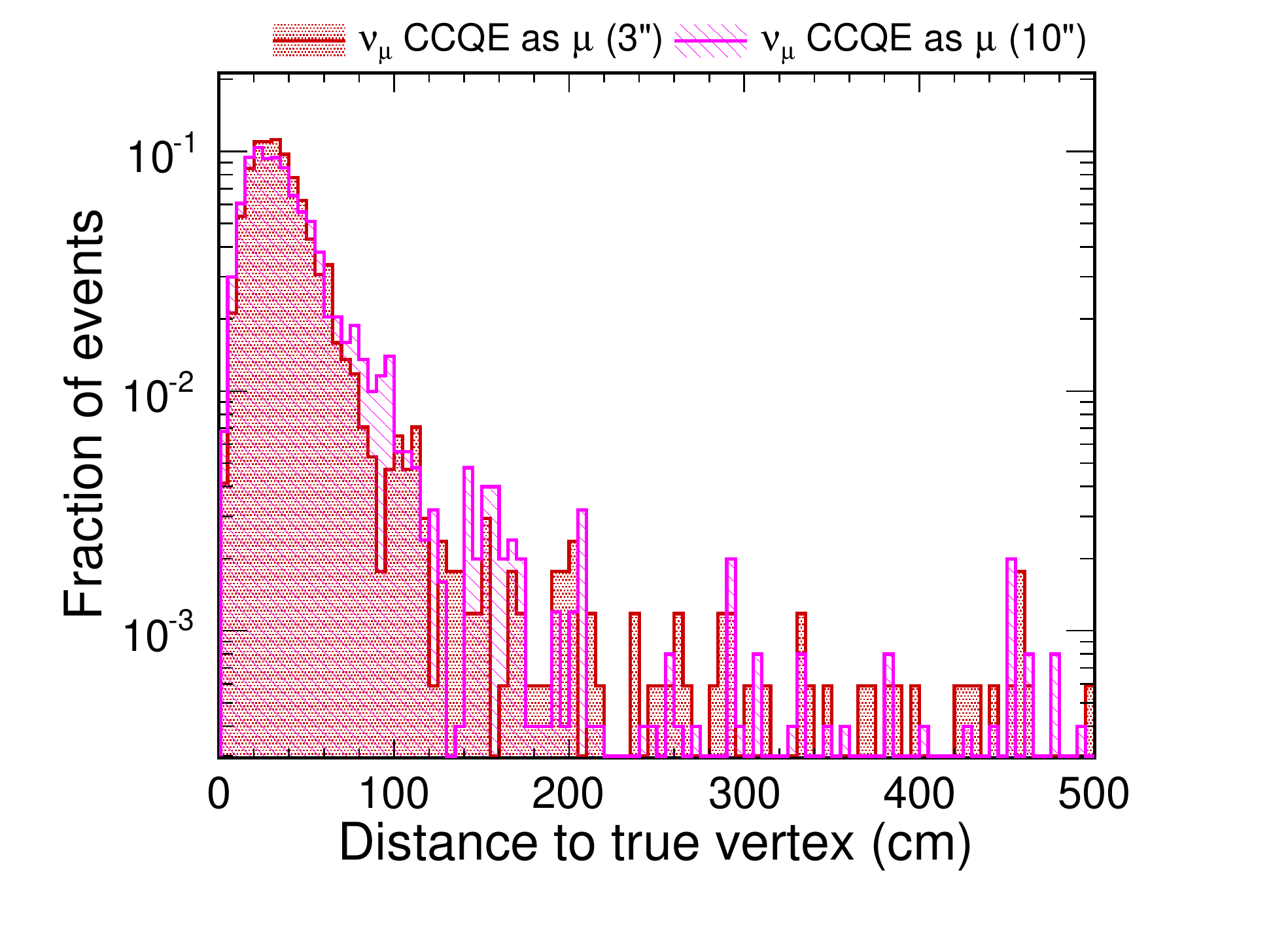} &
      \includegraphics[scale=0.28]{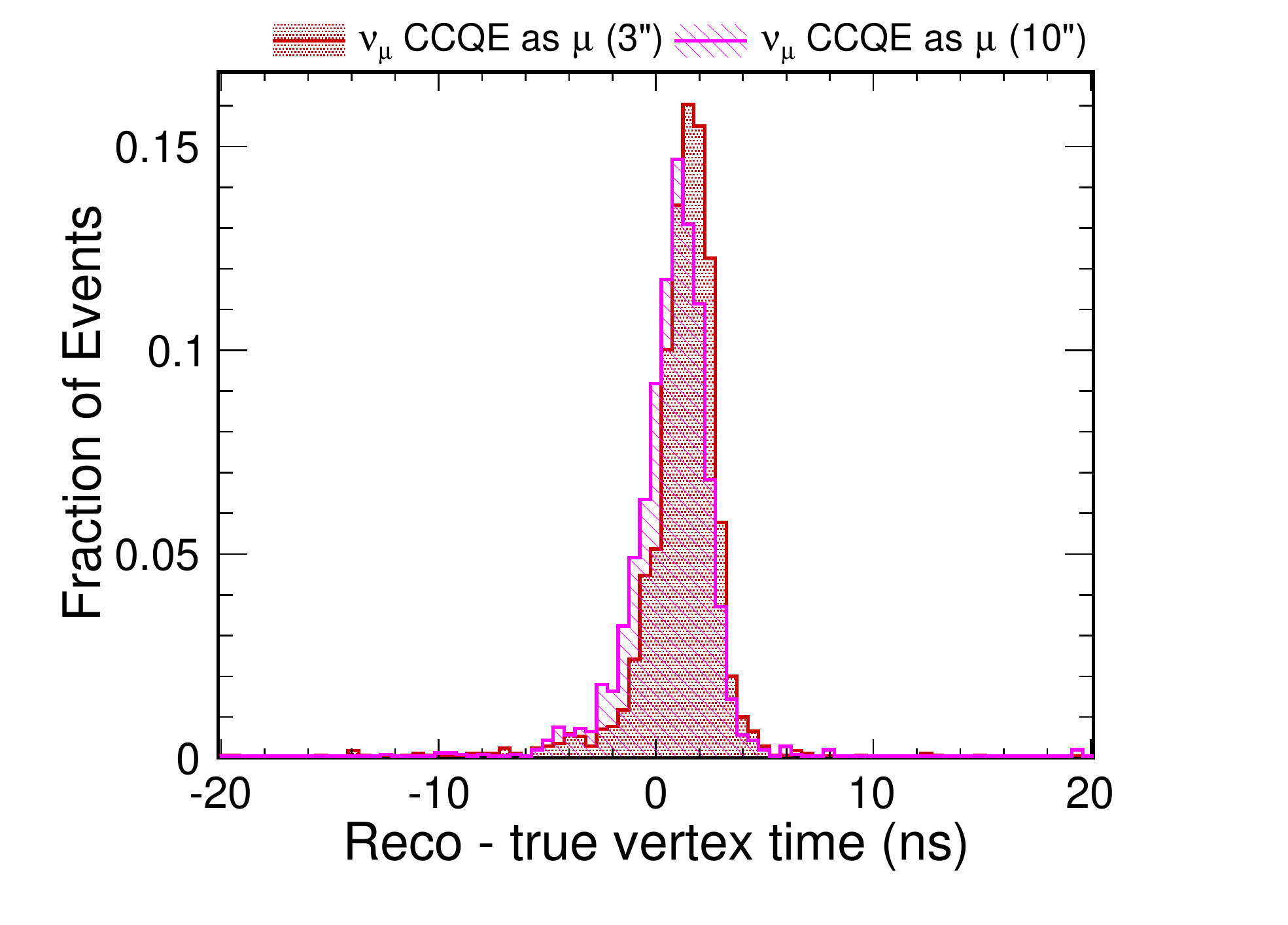} \\
      \includegraphics[scale=0.28]{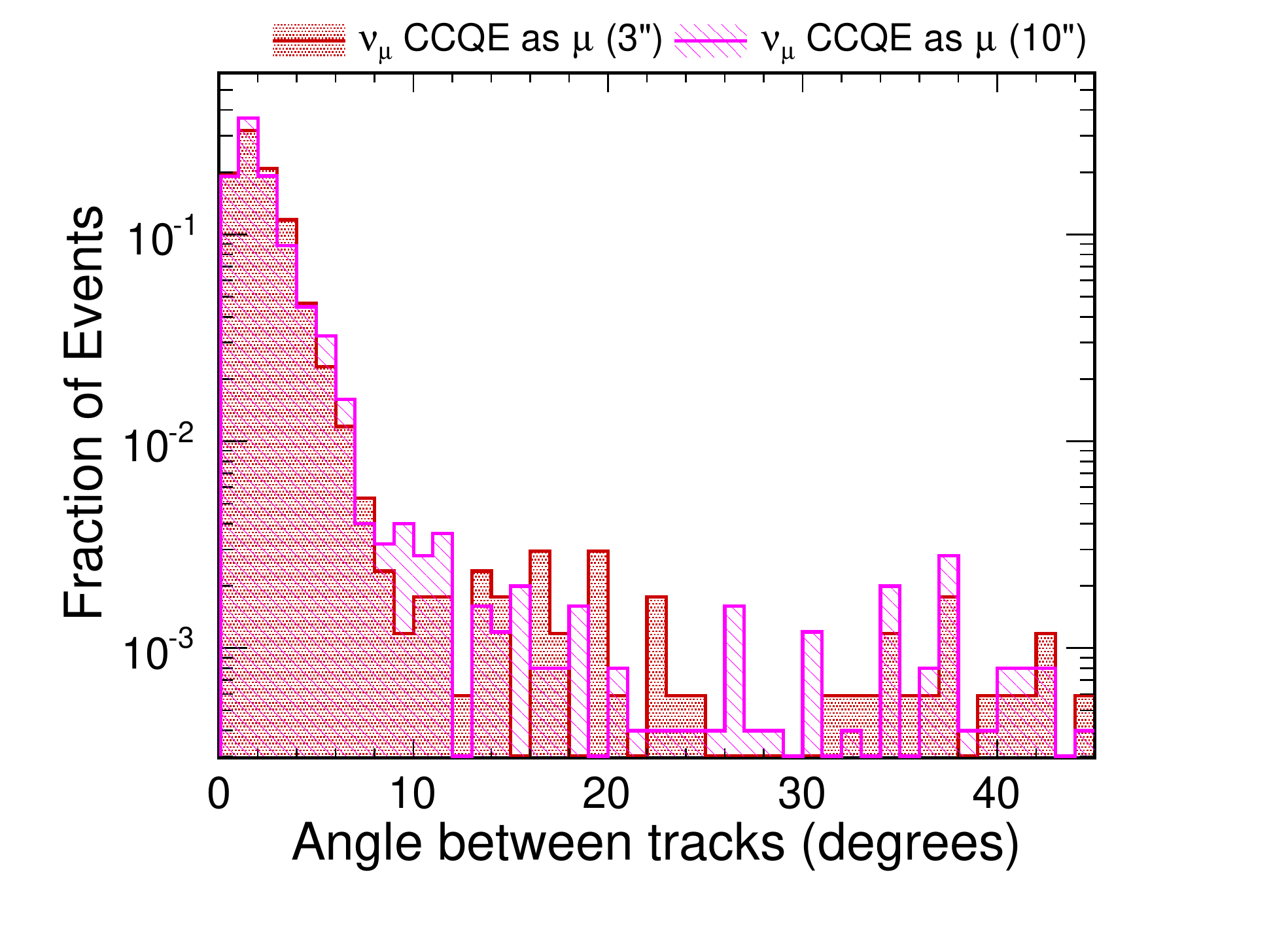} &
      \includegraphics[scale=0.28]{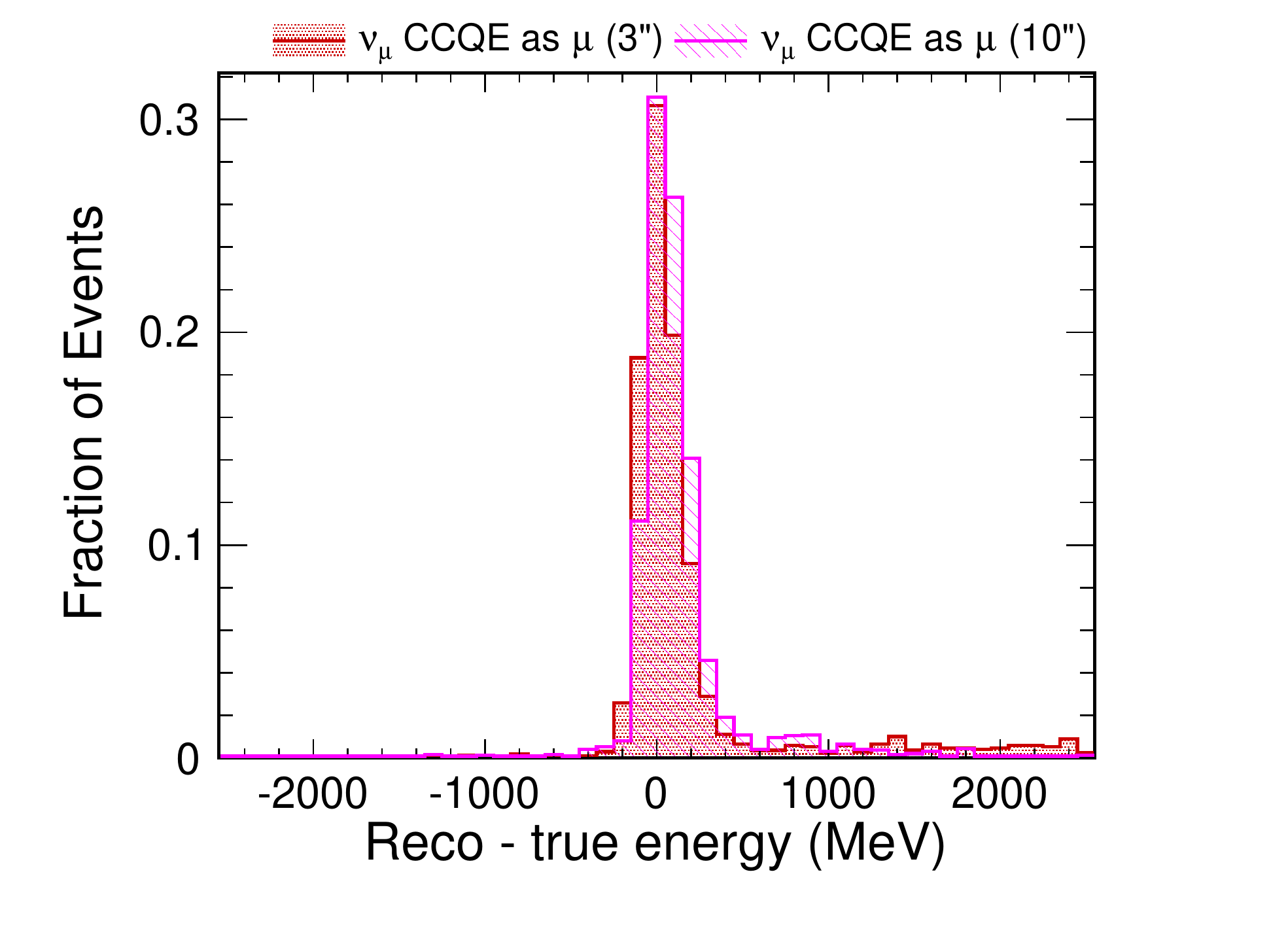} 
    \end{tabular}
  \caption{\label{perf:fig:10vs3NuMu}A comparison of the reconstruction performance for a single track fit to CCQE \numu{} interactions for 10\,inch and 3\,inch PMTs for a photocathode coverage of 10\%. The distributions show (clockwise from top left): the distance of the reconstructed vertex from the simulated vertex, the difference between the reconstructed and simulated values for the vertex time, the difference between the reconstructed and simulated values for the muon energy, and the angle between the reconstructed and simulated track directions.}
  \end{center}
\end{figure}

\begin{figure}
  \begin{center}
    \begin{tabular}{cc}
      \includegraphics[scale=0.28]{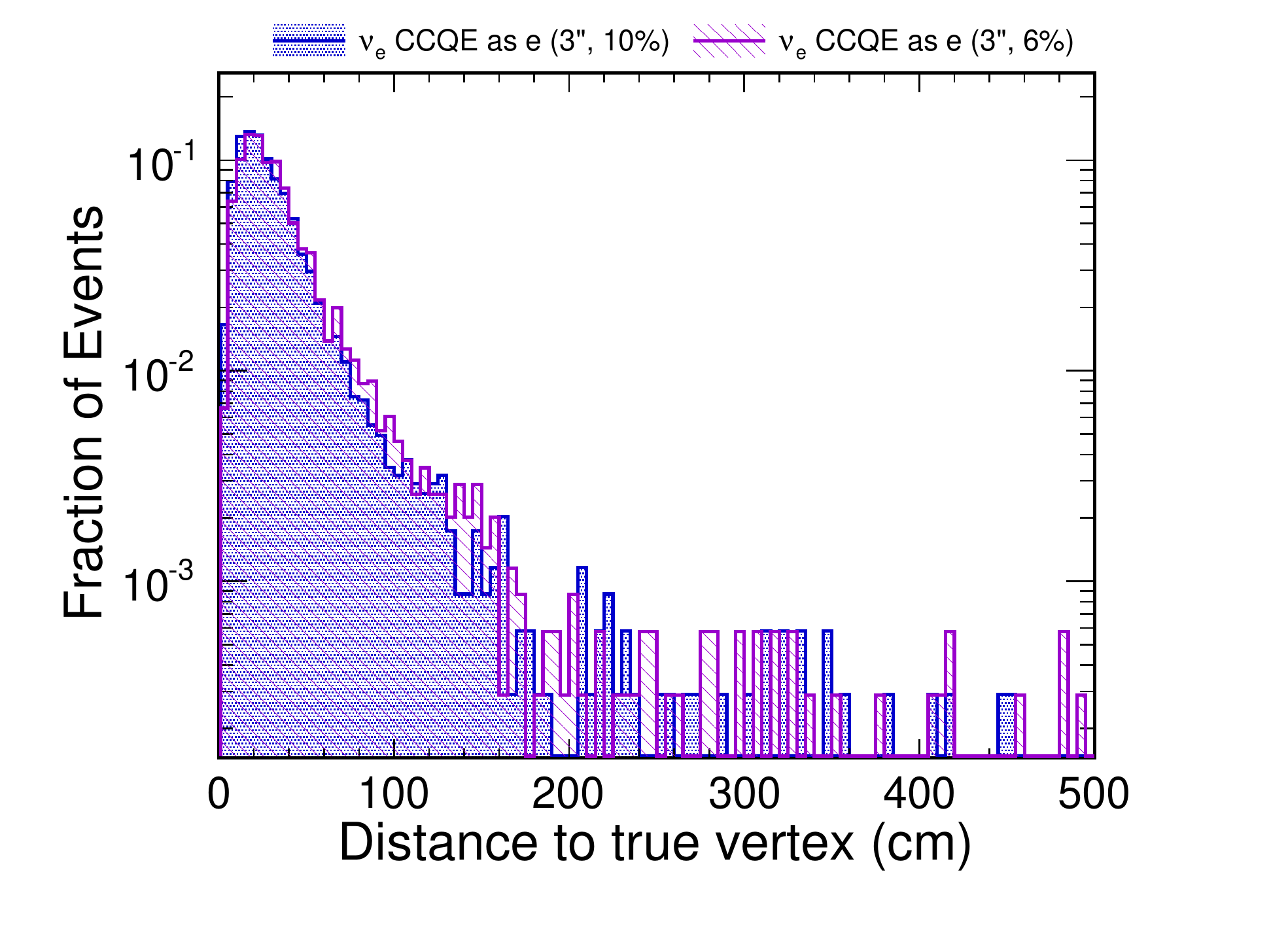} &
      \includegraphics[scale=0.28]{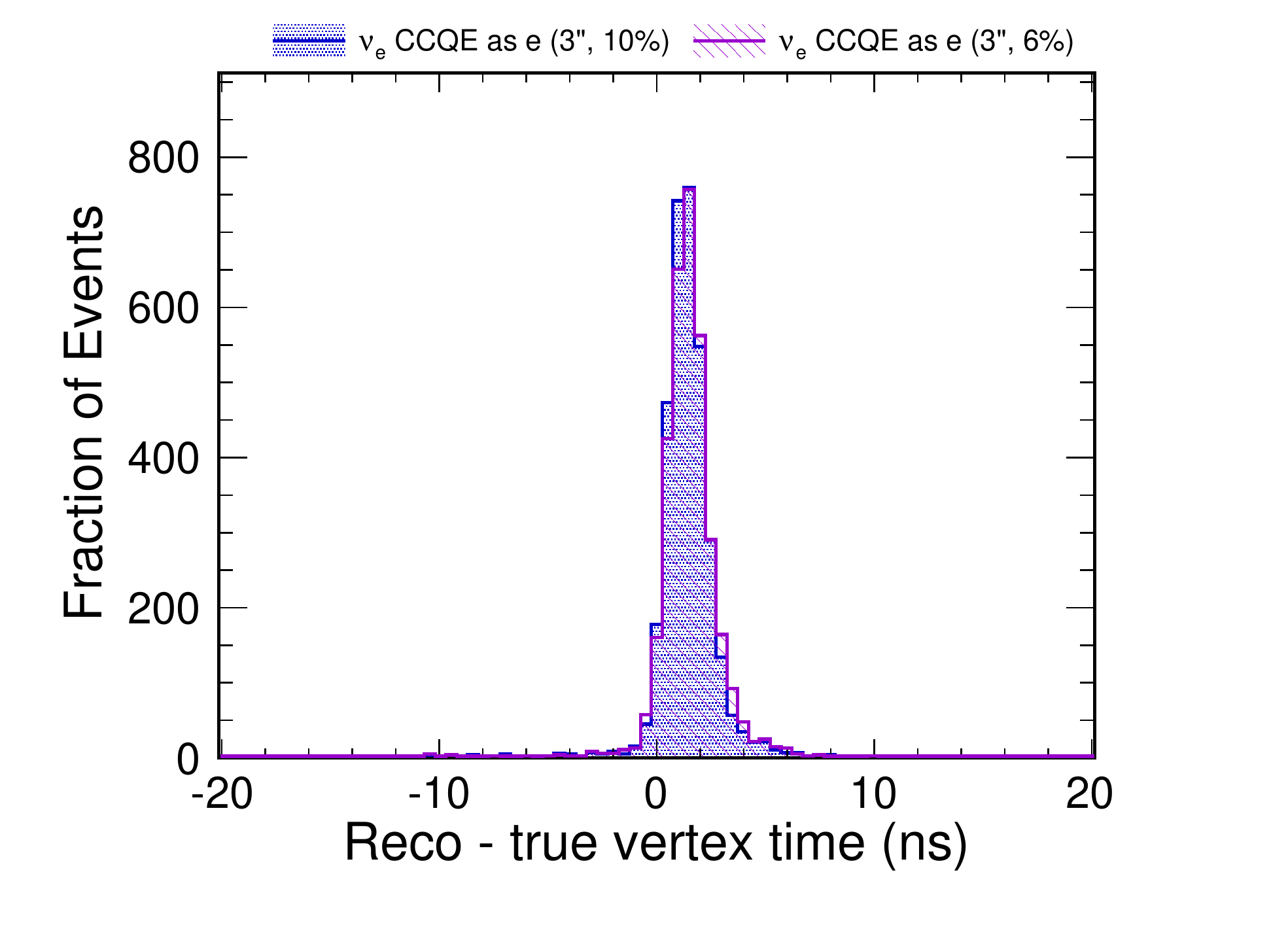} \\
      \includegraphics[scale=0.28]{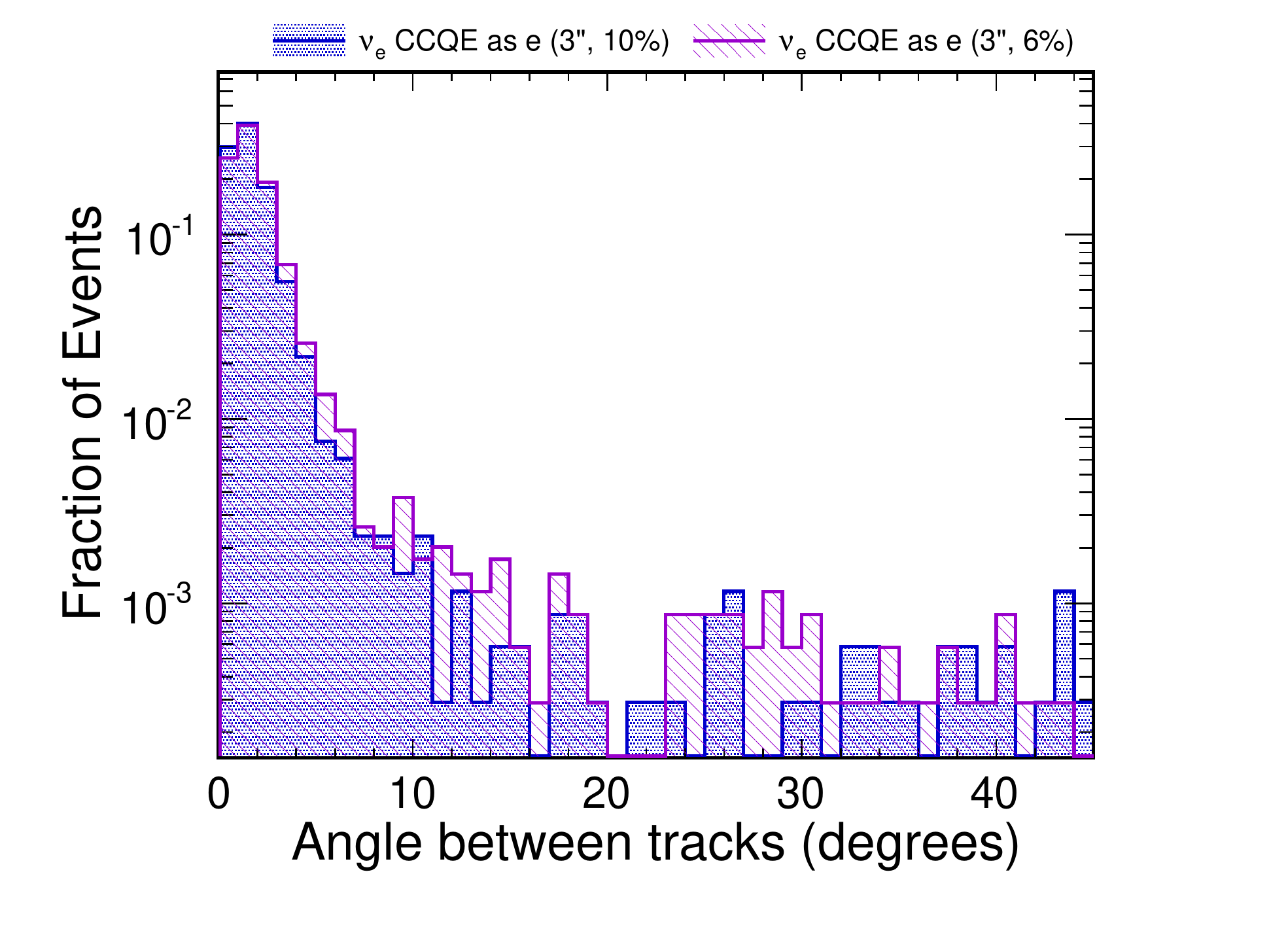} &
      \includegraphics[scale=0.28]{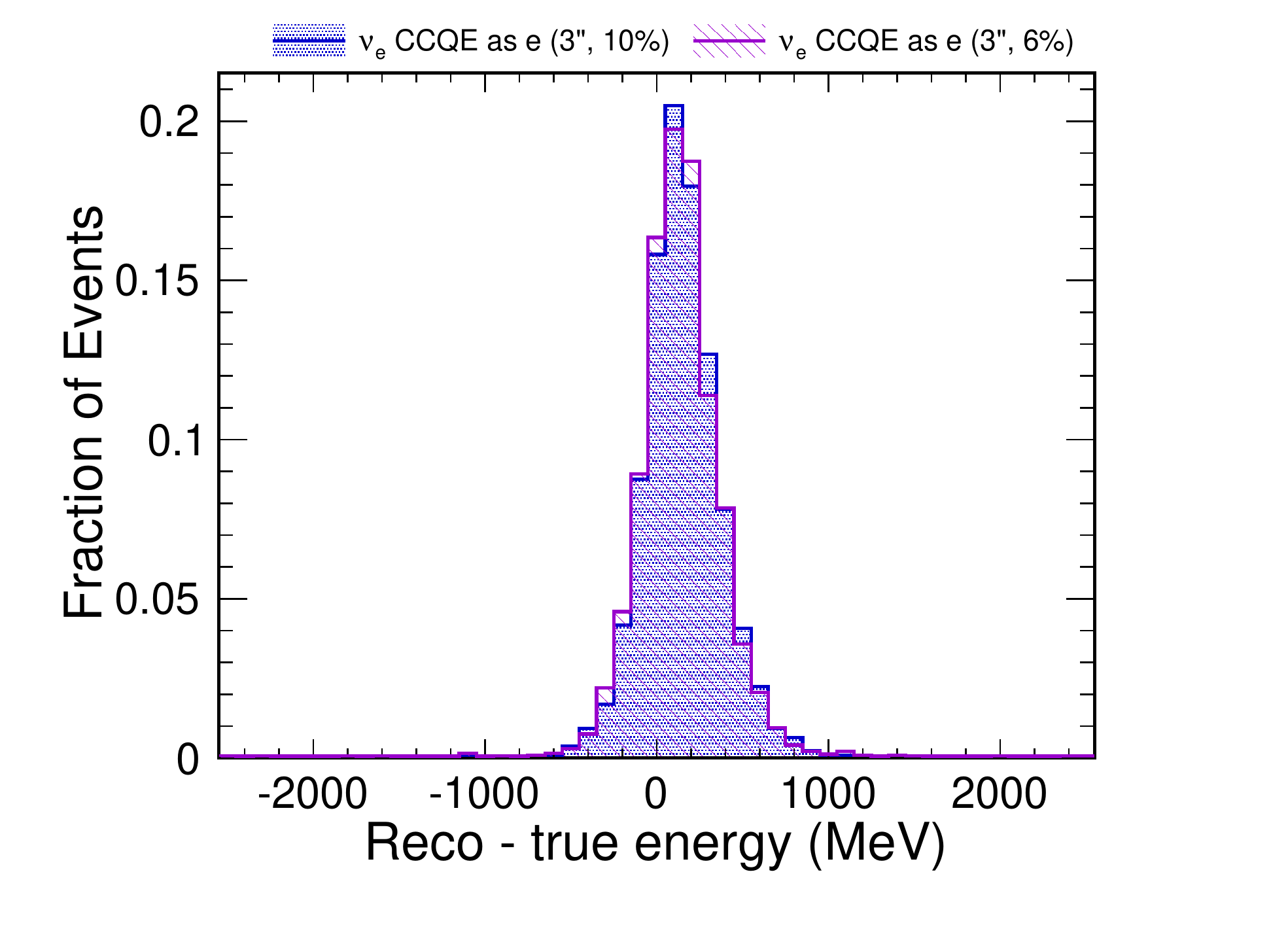} 
    \end{tabular}
  \caption{\label{perf:fig:10vs6NuE}A comparison of the reconstruction performance for a single track fit to CCQE \nue{} interactions for 10\% and 6\% photocathode coverage of 3\,inch PMTs. The distributions show (clockwise from top left): the distance of the reconstructed vertex from the simulated vertex, the difference between the reconstructed and simulated values for the vertex time, the difference between the reconstructed and simulated values for the electron energy, and the angle between the reconstructed and simulated track directions.}
  \end{center}
\end{figure}

\begin{figure}
  \begin{center}
    \begin{tabular}{cc}
      \includegraphics[scale=0.28]{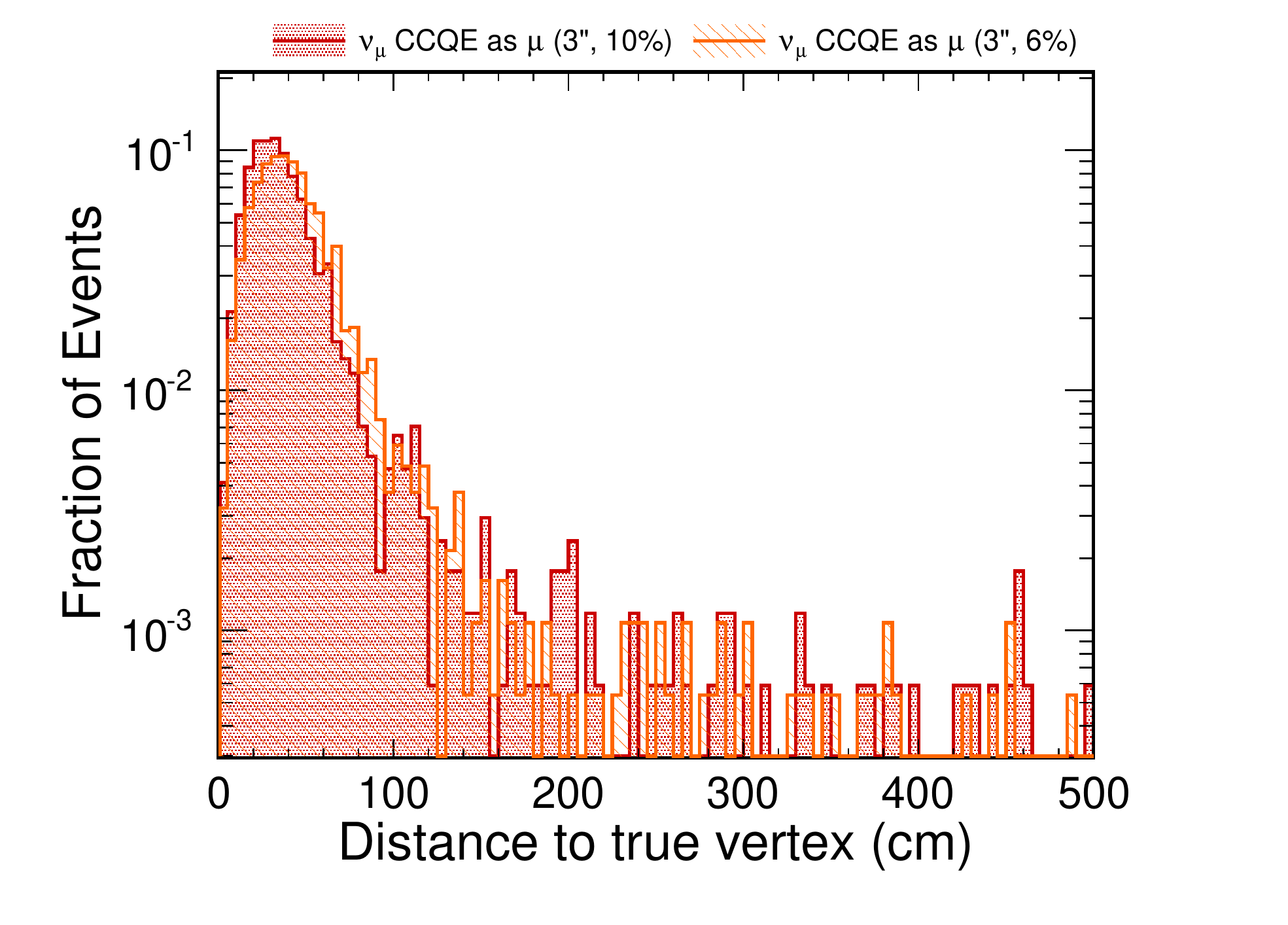} &
      \includegraphics[scale=0.28]{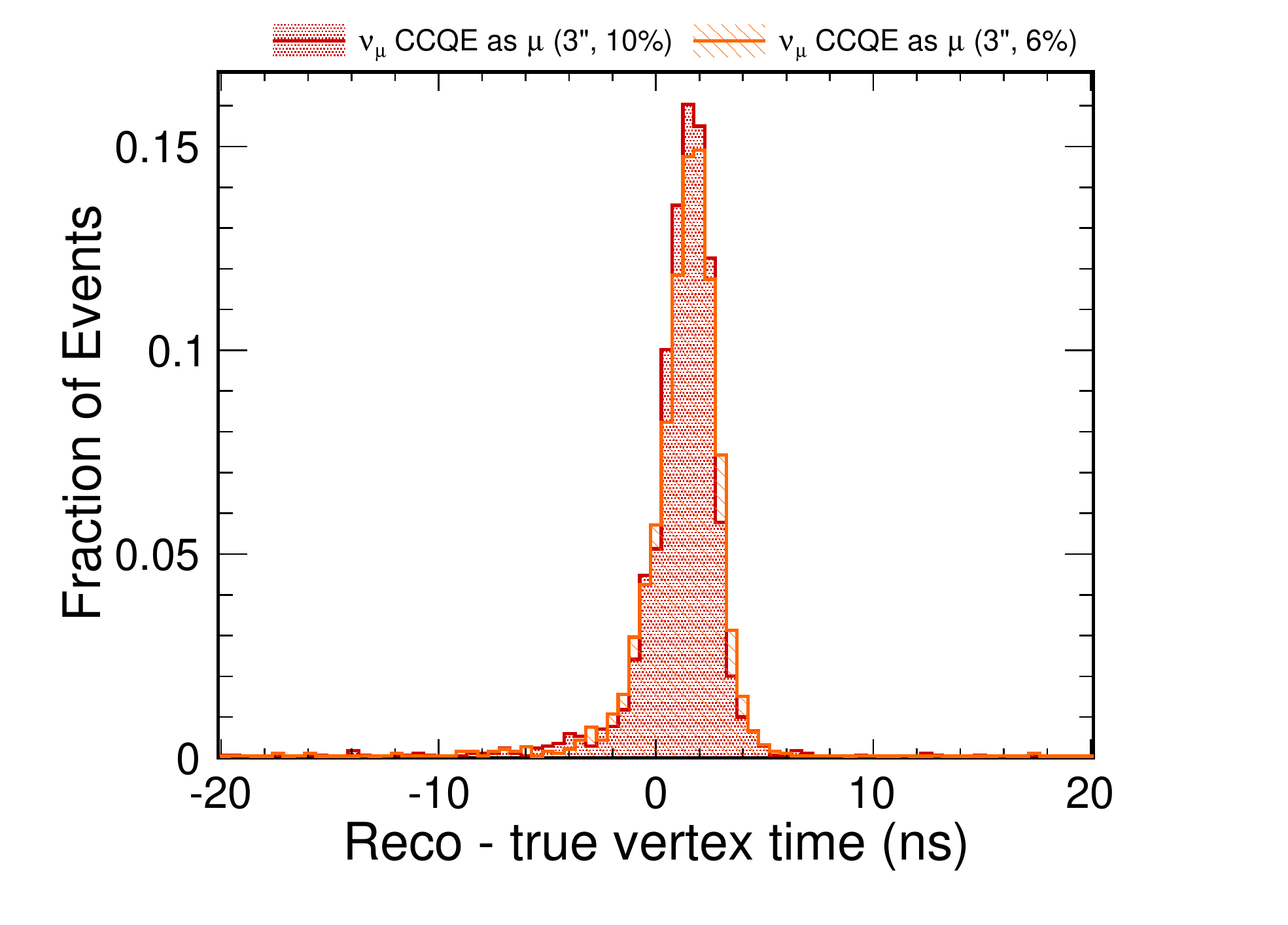} \\
      \includegraphics[scale=0.28]{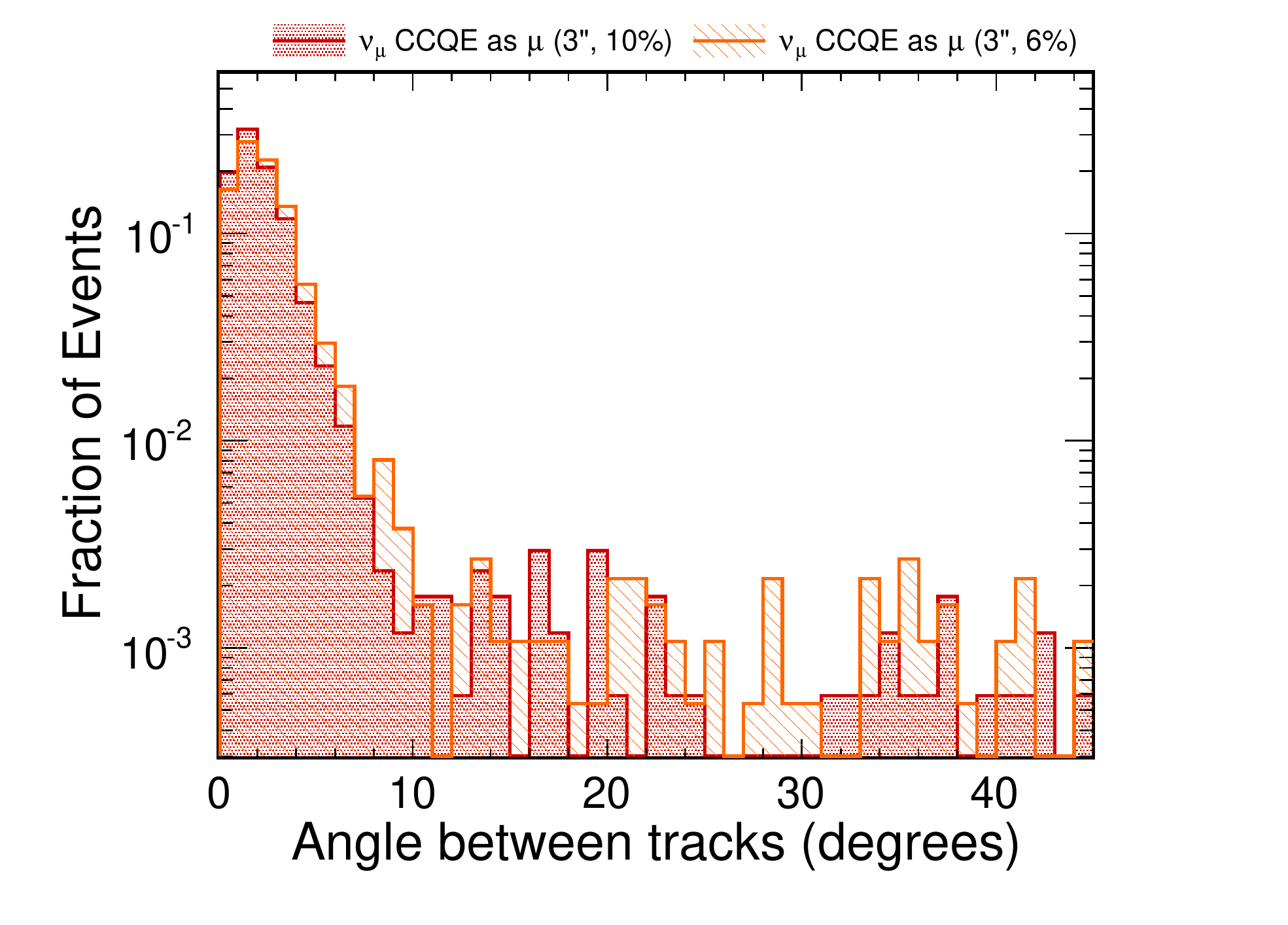} &
      \includegraphics[scale=0.28]{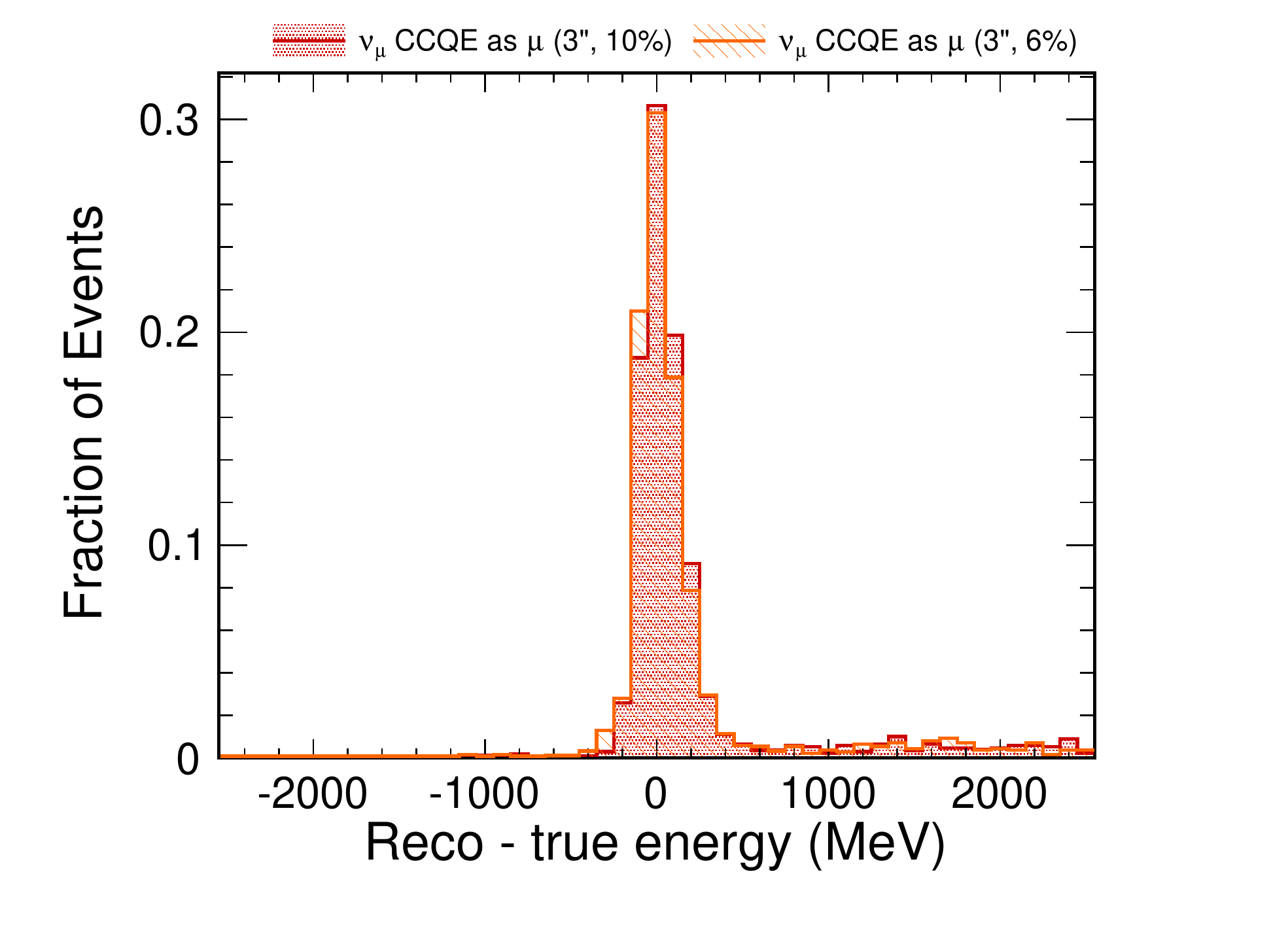}       
    \end{tabular}
  \caption{\label{perf:fig:10vs6NuMu}A comparison of the reconstruction performance for a single track fit to CCQE \numu{} interactions for 10\% and 6\% photocathode coverage of 3\,inch PMTs. The distributions show (clockwise from top left): the distance of the reconstructed vertex from the simulated vertex, the difference between the reconstructed and simulated values for the vertex time, the difference between the reconstructed and simulated values for the muon energy, and the angle between the reconstructed and simulated track directions.}
  \end{center}
\end{figure}

The difference between the reconstructed and simulated distributions for the vertex position, the vertex time, track direction and track energy are shown for CCQE \nue{} and CCQE \numu{} interactions comparing 10\,inch and 3\,inch PMTs, both at 10\% coverage, in Figs. \ref{perf:fig:10vs3NuE} and \ref{perf:fig:10vs3NuMu}, respectively. Similar comparisons are shown in Figs. \ref{perf:fig:10vs6NuE} and \ref{perf:fig:10vs6NuMu} for the 3\,inch PMTs with 10\% and 6\% photocathode coverage for CCQE \nue{} and CCQE \numu{} interactions, respectively. A summary of the data displayed in the figures is given in Table \ref{perf:tab:resolution}. The tables show the resolutions for each of the variables for the three different geometry options, where the resolution for time and energy is given by the width of the Gaussian fit, and by the region containing 68\% of entries for the vertex position and direction since they are non-Gaussian. A small bias in the time and energy variables can be seen, and further studies are on-going to investigate the cause of these shifts. Since the shifts only change the mean of the distributions a correction function could be developed to account for the effect. The results show that for a fixed photocathode coverage of 10\% the 3\,inch PMTs outperform the 10\,inch PMTs. This is expected as the more numerous 3\,inch PMTs provide extra position, timing and shape information. Furthermore, it shows only a small degradation in performance for the 6\% coverage case, which in terms of PMT cost represents a $40\%$ saving. This 6\% coverage with 3\,inch PMTs is the default design for CHIPS-10, and the performance shown suggests that this is a realistic option to use. The studies detailed in the following sections were hence performed using the 3\,inch PMTs with 6\% coverage geometry option.


\begin{table}
  \begin{center}
    \begin{tabular}{llcccc}
      \hline \hline
      \multirow{2}{*}{Sample} & \multirow{2}{*}{Geometry} & \multicolumn{4}{c}{Reconstruction Resolution} \\
      & & Vertex Position (cm) & Vertex Time (ns) & Direction ($^{\circ}$) & Energy (MeV) \\ \hline
      \multirow{3}{*}{CCQE \nue{}} & 10\,inch, 10\%& 35 & 0.9 & 2.1 & 208 \\
      & 3\,inch, 10\%& 35 & 0.84 & 1.9 & 210 \\
      & 3\,inch, 6\%& 38 & 0.89 & 2.1 & 211 \\
      \multirow{3}{*}{CCQE \numu{}} & 10\,inch, 10\%& 47 & 1.35 & 2.6 & 113 \\
      & 3\,inch, 10\%& 44 & 1.14 & 2.7 & 110 \\
      & 3\,inch, 6\%& 51 & 1.28 & 3.0 & 113 \\ \hline \hline
    \end{tabular}
  \caption{\label{perf:tab:resolution}The resolutions of various parameters from a single electron (muon) track fit to a sample of CCQE \nue{} (\numu{}) interactions with energies following those expected from the \numi{} beam. The resolutions were calculated from the width of the distributions shown in Figures \ref{perf:fig:10vs3NuE} - \ref{perf:fig:10vs6NuMu}.}
  \end{center}
\end{table}

\section{Particle Identification}
The fact that the method is based on likelihoods naturally allows for the use of likelihood ratios as particle identification (PID) variables. 
However, it was found that other variables in addition to the likelihoods have some power to separate event types. In addition, there is more power to be extracted from the likelihoods by considering the charge and time components separately, as opposed to simply adding the two components together. 

For this reason, two artificial neutral networks (ANNs) were developed using the TMVA package \cite{TMVA} within the ROOT framework \cite{root}. The first of the networks is used to perform a separation of electron-like and muon-like interactions and is henceforth referred to as \annem{}. The second network, called \annnc{}, is designed to to discriminate between electron-like and NC interactions.  The list of variables used in the networks is as follows:
\begin{itemize}
  \item $\Delta\ln\mathcal{L}$ between muon and electron hypotheses for time and charge components
  \item $N_{\textrm{hits}}$, the number of hit PMTs
  \item $\frac{\Delta\ln\mathcal{L}_{\textrm{charge}}}{N_{\textrm{hits}}}$
  \item Fraction of hits outside, within, and inside the central hole of the ring for both muon and electron hypotheses
  \item Fraction of predicted charge outside the ring for both muon and electron hypotheses
  \item Ratio of the total predicted charge to the total measured charge for both muon and electron hypotheses
  \item The ratio of the reconstructed energy to the total measured charge
  \item \annnc{} only: reconstructed track direction under the electron hypothesis
  \item \annnc{} only: fraction of hits in the downstream half of the detector
\end{itemize}
The two \annnc{} only variables reflect the fact that NC events are typically less collimated along the neutrino beam axis, but do not provide any information for the other ANN. The two networks each produce a single output parameter that can be used to discriminate between the different hypotheses. These outputs are referred to as $a_{e\mu}$ and $a_{NC}$ for \annem{} and \annnc{}, respectively. The variables are distributed in such a way that a value close to one suggests that the event is signal-like, where the signal in both cases is electron-like, and a value close to zero suggests the event is background-like.

A sample of CCQE \nue{} and CCQE \numu{} interactions was used to train the \annem{} classifier so that the ANN was trained on the cleanest subset of CC interactions. Similarly, the \annnc{} classifier was trained on a sample of CCQE \nue{} and NC interactions. The same preselection described in Section \ref{sec:pmtComp} was applied to both the training and testing samples. An additional requirement on the \annnc{} training samples was that the output variable from \annem{}, $a_{e\mu}$ is greater than 0.8. This additional requirement was to ensure that the \annnc{} network did not try to predominantly remove CC \numu-like events.

\subsection{Particle identification performance}
In order to study the PID performance, samples of CC \nue{}, CC \numu{} and NC $\nu$ interactions were processed through the reconstruction algorithm twice, each time using a single track fit. On the first pass they were all fitted under the electron track hypothesis, and on the second pass they were fitted assuming the muon hypothesis. Fitting each event under both hypotheses provides the basis to distinguish which hypothesis is the more likely.

\begin{figure}
  \centering
  \includegraphics[scale=0.5]{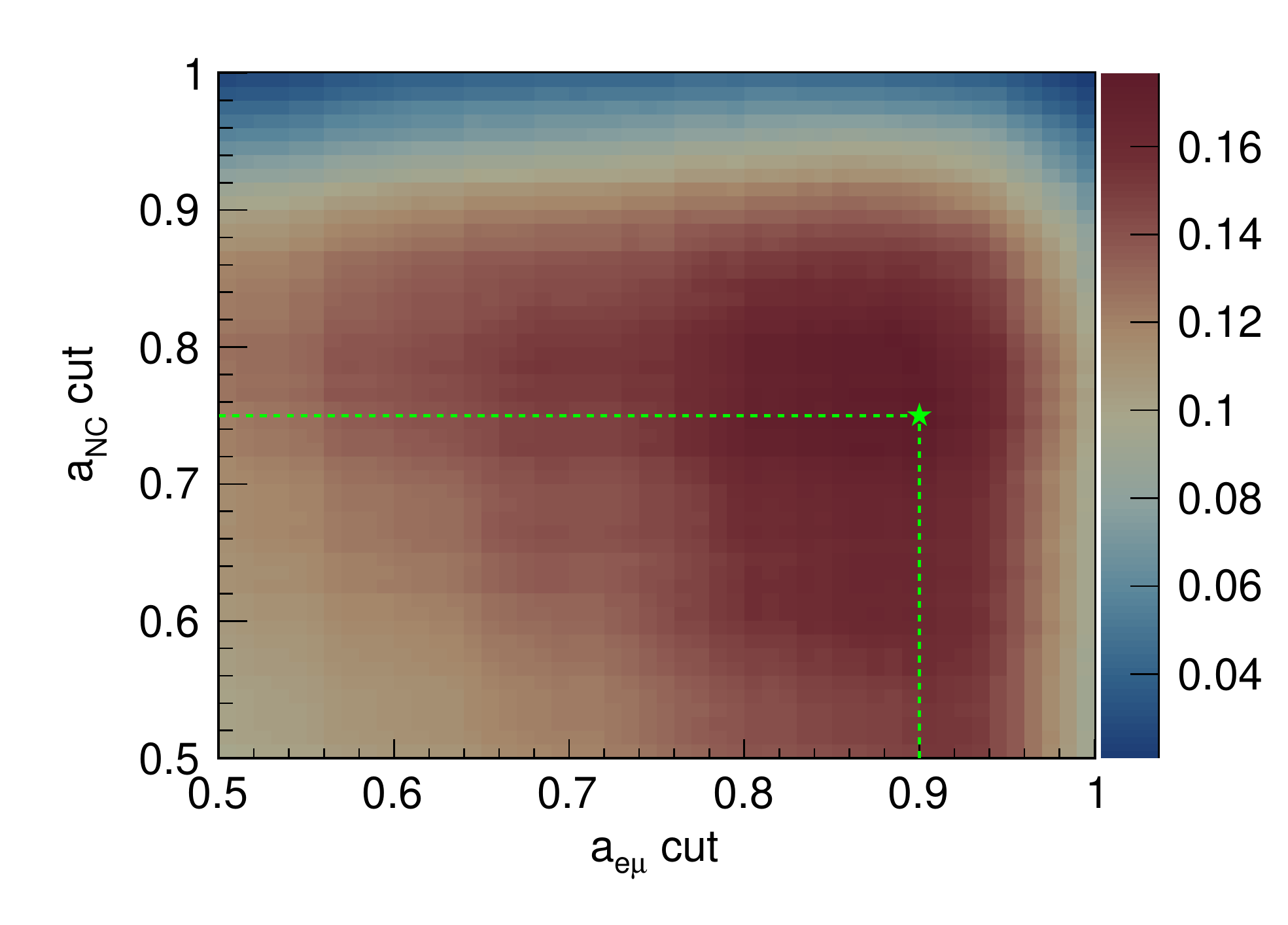}
  \caption{\label{perf:fig:ep}The value of CCQE \nue{} efficiency $\times$ purity (colour axis) as a function of the cut value on the output variables of \annem{} and \annnc{}. The green star shows the optimised cut position that maximises the figure of merit, with those events falling in the upper right region selected as CC \nue{} candidates.}
\end{figure}

The value $\epsilon\mathbb{P}$, where $\epsilon$ is the CC \nue{} selection efficiency and $\mathbb{P}$ is the CC \nue{} selection purity, was calculated over a grid of cut values on the outputs of the \annem{} and \annnc{} classifiers and is shown in Fig. \ref{perf:fig:ep}. The cut values that maximise the figure of merit are shown as the green star such that the requirements for an event to be selected as a \nue{} candidate event are $a_{e\mu} > 0.9$ and $a_{NC} > 0.75$.

\begin{figure}
  \centering
  \includegraphics[scale=0.40]{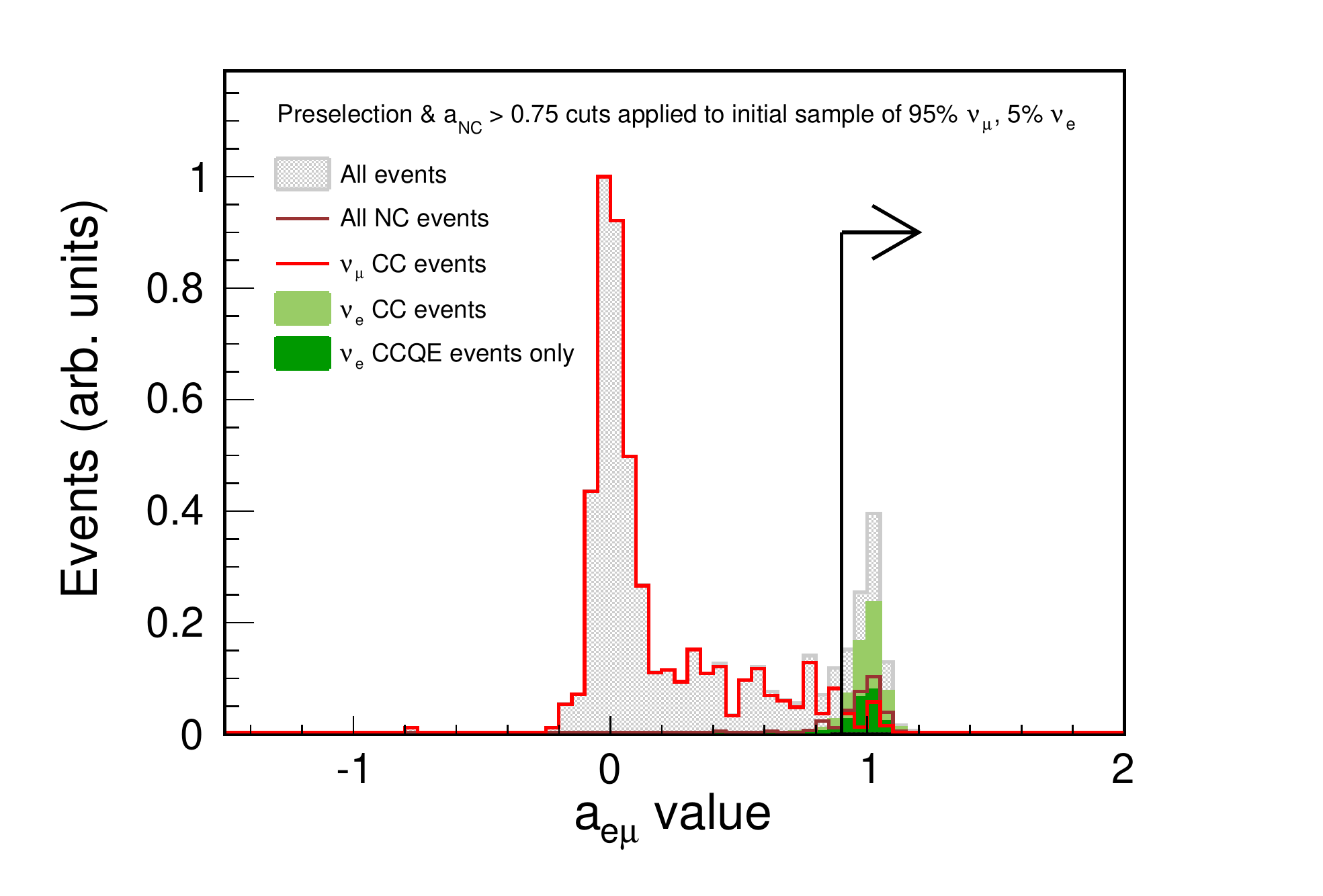}
  \includegraphics[scale=0.40]{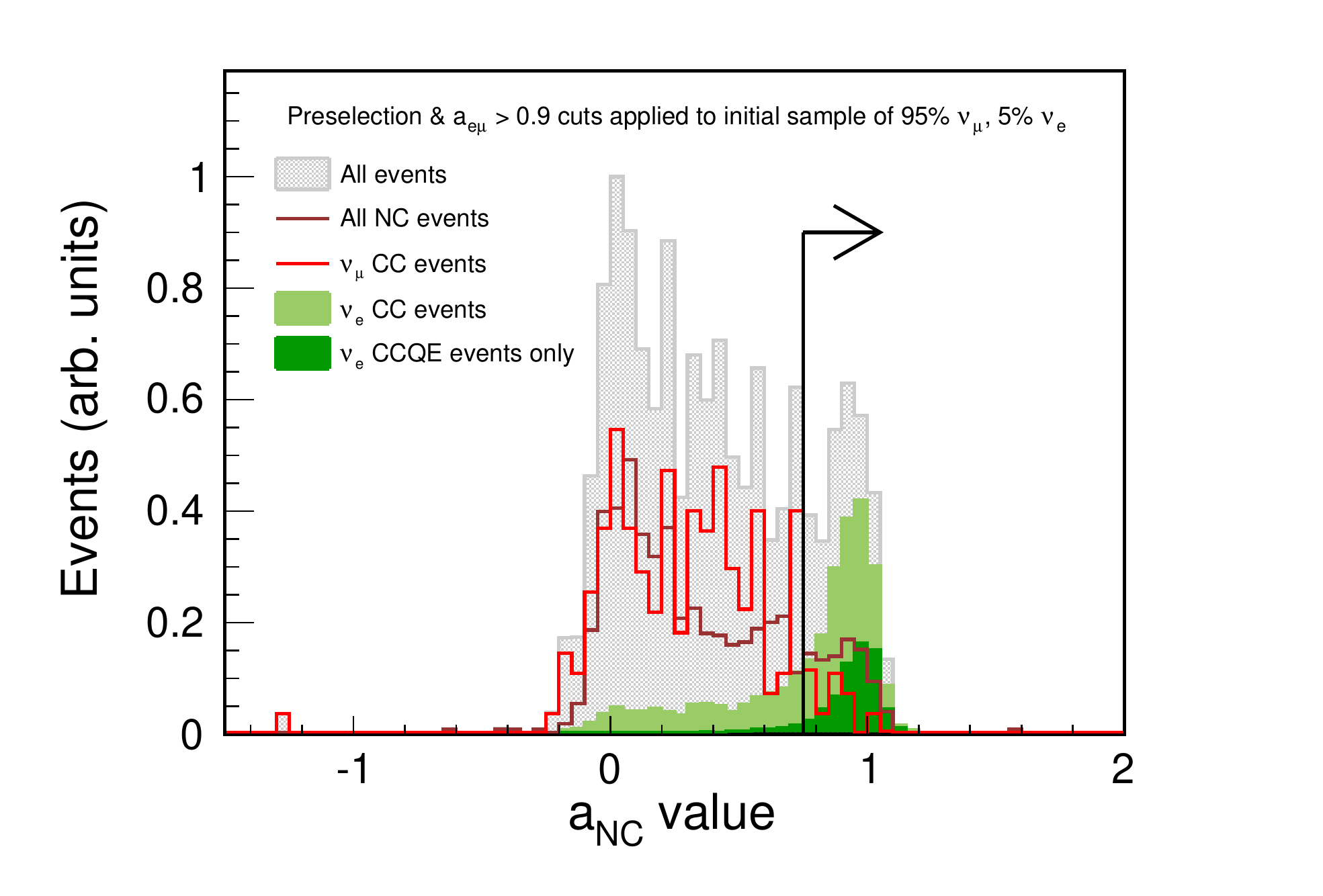}
  \caption{\label{perf:fig:ANNvars}The output distribution from each of the ANNs, shown after the application of the preselection and the optimised cut on the other ANN output, for CC \nue{}, CCQE \nue{}, CC \numu{} and NC interactions. The cut on the other ANN was applied in order to show more clearly the events that each network was designed to separate. The black line and arrow shows the optimised cut value on the displayed network output variable used to selected \nue{} events.}
\end{figure}

The ANNs were trained to look for CCQE \nue{} interactions, but in the CHIPS neutrino spectrum CCQE interactions only make up approximately 20\% of the total CC sample for both \numu{} and \nue{} flavours. It is therefore important to test the performance of these ANNs on the total expected event samples that also include CC RES and CC DIS interactions, in addition to the CCQE and NC events.  Figure \ref{perf:fig:ANNvars} shows the distribution of $a_{e\mu}$ (left) and $a_{NC}$ (right), including the preselection and a cut on the other ANN, for CC \numu{}, CC \nue{} (and the CCQE subset) and NC interactions. The relative number of events in each sample is set to match the predictions for the beam composition at the CHIPS-10 detector. A comparison of the shapes of the two ANN output variables shows that it is relatively simple to separate the CC \numu{} and CC \nue{} interactions, but the NC background is more difficult to remove. It also shows that whilst the networks were trained on CCQE interactions, even a single ring fit can do well at identifying more complex CC \nue{} interactions in order to perform a counting experiment.

\begin{figure}
  \centering
  \includegraphics[scale=0.5]{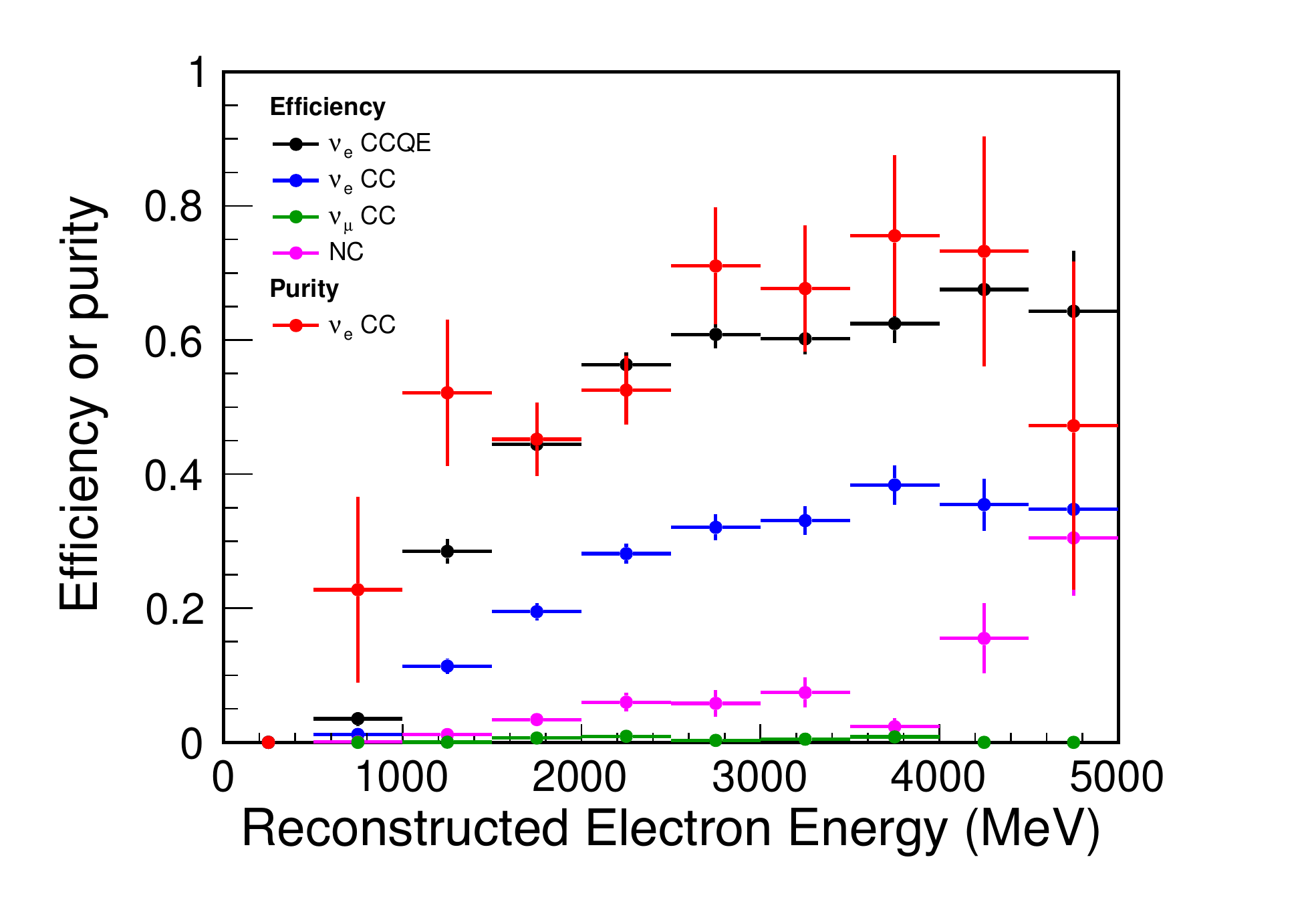}
  \caption{\label{perf:fig:effPur}The efficiency and purity of various components with a selection requiring $a_{e\mu} > 0.9$ and $a_{NC} > 0.75$ in addition to the preselection, shown as a function of the reconstructed energy under the electron hypothesis.}
\end{figure}

A summary of the selection efficiency and purity is shown in Fig. \ref{perf:fig:effPur} for the requirement that each event must have $a_{e\mu} > 0.9$ and $a_{NC} > 0.75$ in addition to passing the preselection. In particular, the efficiency is plotted for the signal CC \nue{} interactions, and also for the CCQE \nue{} subsample. Additionally, the efficiencies for the CC \numu{} and NC backgrounds are also shown. This shows that even performing a simple single track fit, a sample of CC \nue{} interactions can be selected with approximately 30\% efficiency and 58\% purity. The background rejection for CC \numu{} and NC interactions is 99.6\% and 98.3\%, respectively.  The comparison of the CCQE \nue{} efficiency and the CC \nue{} efficiency, shown in black and blue respectively, shows that the selectors still work well even for the more complex event topologies.

\subsubsection{Fitting $\pi^{0}$ events}
\begin{figure}
  \centering
  \includegraphics[scale=0.4]{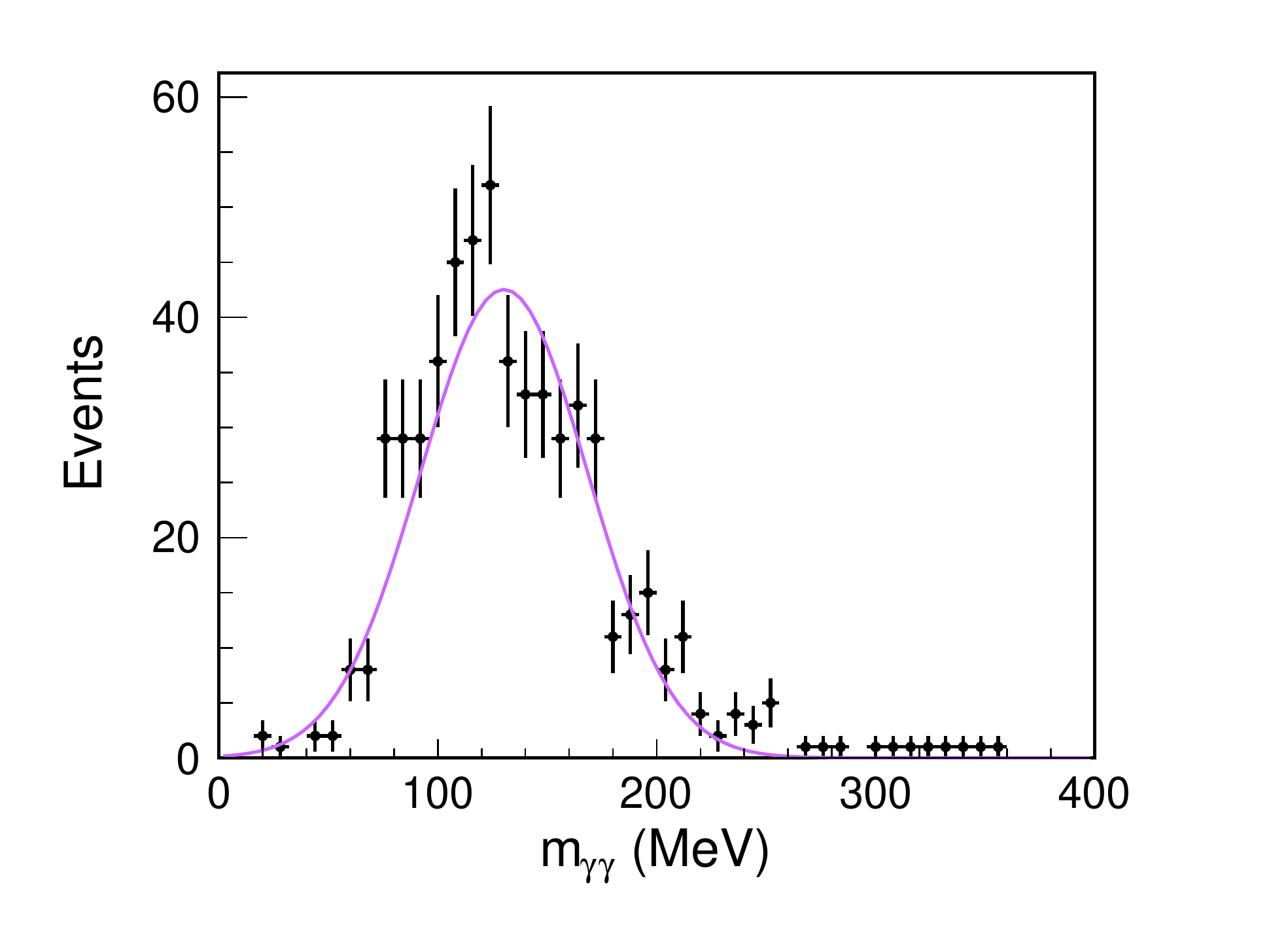}
  \caption{\label{perf:fig:invMass}The invariant mass of the two photon tracks reconstructed using the two track $\pi^{0}$ fit for all events where the two decay photons were reconstructed with energies greater than 150\,MeV.}
\end{figure}

A sample of simulated $\pi^{0}$ mesons were obtained by extracting them from the standard GENIE NC interactions to ensure a representative energy distribution, and were then passed through the simulation and reconstruction chain. The events were fit under the $\pi^{0}$ hypothesis without constraining the photon energies and produced the invariant mass distribution shown in Fig. \ref{perf:fig:invMass}, with the requirement that each reconstructed photon had an energy greater than $150\,$MeV. A clear peak can be seen in the distribution, and a Gaussian fit yields a mean of $130\pm34\,$MeV in excellent agreement with the $\pi^{0}$ mass $m_{\pi^{0}} = 134.98\,$MeV~\cite{pdg}. It shows that the fitter is capable of reconstructing $\pi^{0}$ mesons of the energy expected in the CHIPS-10 detector even with only a 6\% coverage of 3 inch PMTs.

The $\pi^{0}$ fitter will be integrated with the rest of the PID algorithms in order to help further reduce the NC background to the \nue{} appearance search.

\section{Conclusions}
A series of event reconstruction algorithms have been developed for the CHIPS experiment, using charge and time information from all PMTs. The charge component was based on the method developed by MiniBooNE, whereas the time component was developed from first principles. The software was used to perform a comparison between three geometry options and showed that a very low 6\% photocathode coverage of 3\,inch PMTs is a plausible option for CHIPS-10. The low photocathode coverage of small PMTs is a vital part of minimising the cost of the experiment and hence the demonstrated ability to successfully reconstruct neutrino interactions under these conditions is a very important result.

A simple single track fitting technique under both the electron and muon hypothesis was tested on a representative sample of events predicted for CHIPS-10 as exposed by the \numi{} beam. Signal CC \nue{} events were selected with 30\% efficiency and 58\% purity. The particle identification was performed using two neural networks: \annem{}, trained to separate CCQE \nue{} events from CCQE \numu{} interactions; and \annnc{} designed to separate CCQE \nue{} interactions from NC interactions. The selection efficiency and background rejection compare well to the values assumed in the CHIPS Letter of Intent \cite{chipsLOI}.

The reconstruction software and particle identification capability will be further studied in the future in order to optimise the physics capability of the detector. Firstly, a $\pi^{0}$ fitter has been demonstrated to work on a sample of $\pi^{0}$ interactions with representative energies of those expected to be produced in NC interactions in the NuMI beam. The $\pi^{0}$ fitter will be integrated with the two PID ANNs in order to minimise the background arising from NC interactions as well as those from CC \numu{} events. Secondly, a technique has been developed in order to fit cosmic-ray muon interactions that overlap with the beam interactions. Further studies will determine the effects of the cosmic-ray muons on the reconstruction of the beam interaction and determine the level of degradation of the various reconstruction resolutions. Finally, additional methods of performing particle identification will be investigated. For example, a Convolutional Neural Network based event classifier using minimal event reconstruction has been shown to give a 30\% improvement in the \nue{} event selection efficiency on the NO$\nu$A experiment~\cite{novaCNN}. This technique draws from image feature recognition, which should also be applicable to a water Cherenkov detector.

\section{Acknowledgments}
This work was supported in the UK by the Leverhulme Trust.

\section{Bibliography}
\bibliography{chipsBib}

\end{document}